%% file: Kumar_Elia_Jalden_LRaidedLD_Arxiv.tex
\documentclass[10pt,twocolumn,twoside]{IEEEtran}
\usepackage{calc}

\usepackage{multirow}
\usepackage{cite}
\usepackage{graphicx,subfigure}
\usepackage{psfrag}
\usepackage{amsmath,amssymb}
\usepackage{color}

\input defs.tex

\input macros.tex

\def\nt{n_\mathrm{T}}
\def\nr{n_\mathrm{R}}

\def\ints{\mathbb{Z}}

\def\kron{\otimes}
\def\fro{_\mathrm{F}}
\def\kron{\otimes}

\def\C{\mathbb{C}}

\newtheorem{remark}{Remark}

\begin{document}
\sloppy

\title{Achieving a vanishing SNR-gap to exact lattice decoding at a subexponential complexity}
\author{Arun Singh, Petros Elia and Joakim~Jald\'en
\thanks{The research leading to these results has received funding from the European Research Council under the European Community's Seventh Framework Programme (FP7/2007-2013) / ERC grant agreement no. 228044, from the Swedish Foundation for Strategic Research (SSF) / grant ICA08-0046, from FP7/2007-2013 grant agreement no. 257616 (CONECT), and from the Mitsubishi Electric R$\&$D Centre Europe project Home-eNodeBS.
}
\thanks{A. Singh and P. Elia are with the Mobile Communications Department, EURECOM, Sophia Antipolis, France (email: \{singhak, elia\}@eurecom.fr)}
\thanks{J. Jald\'en is with the ACCESS Linnaeus Center, Signal Processing Lab, KTH Royal Institute of Technology, Stockholm, Sweden (email: jalden@kth.se)}
}

%%%%%%%%%%%%%%%%%%%%%%%%%%%%%%%%%%%%%%%%%%%%%%%%%%%%%%%%%%%%%%%%%%%%%

\maketitle
\thispagestyle{empty}

%%%%%%%%%%%%%%%%%%%%%%%%%%%%%%%%%%%%%%%%%%%%%%%%%%%%%%%%%%%%%%%%%%%%%%%%%%%%%%%%%%

\begin{abstract}

The work identifies the first lattice decoding solution that achieves, in the general outage-limited MIMO setting and in the high-rate and high-SNR limit, both a vanishing gap to the error-performance of the (DMT optimal) exact solution of preprocessed lattice decoding, as well as a computational complexity that is subexponential in the number of codeword bits.  The proposed solution employs lattice reduction (LR)-aided regularized (lattice) sphere decoding and proper timeout policies.
These performance and complexity guarantees hold for most MIMO scenarios, all reasonable fading statistics, all channel dimensions and all full-rate lattice codes.

In sharp contrast to the above very manageable complexity, the complexity of other standard preprocessed lattice decoding solutions is revealed here to be extremely high.  Specifically the work is first to quantify the complexity of these lattice (sphere) decoding solutions and to prove the surprising result that the complexity required to achieve a certain rate-reliability performance, is exponential in the lattice dimensionality and in the number of codeword bits, and it in fact matches, in common scenarios, the complexity of ML-based solutions.
Through this sharp contrast, the work was able to, for the first time, rigorously demonstrate and quantify the pivotal role of lattice reduction as a special complexity reducing ingredient.

Finally the work analytically refines transceiver DMT analysis which generally fails to address potentially massive gaps between theory and practice.  Instead the adopted vanishing gap condition guarantees that the decoder's error curve is arbitrarily close, given a sufficiently high SNR, to the optimal error curve of exact solutions, which is a much stronger condition than DMT optimality which only guarantees an error gap that is subpolynomial in SNR, and can thus be unbounded and generally unacceptable for practical implementations.
\end{abstract}

%%%%%%%%%%%%%%%%%%%%%%%%%%%%%%%%%%%%%%%%%%%%%%%%%%%%%%%%%%%%%%%%%%%%%%%%%%%%%%%%%%%

%%%%%%%%%%%%%%%%%%%%%%%%%%%%%%%%%%%%%%%%%%%%%%%%%%%%%%%%%%%%%%%%%%%%%%%%%%%%%%%%%%%

\section{Introduction}
The work applies to the general setting of outage-limited MIMO communications, where MIMO techniques offer significant advantages in terms of increased throughput and reliability, although at a cost of a potentially much higher computational complexity for decoding at the receivers.
This high complexity brings to the fore the need for efficient decoders that tradeoff error-performance with complexity in a better manner than computationally expensive decoders like the strictly optimal maximum-likelihood (ML) decoder.

Specifically in terms of ML-based decoding, the use of the brute-force ML decoder, introduces a complexity that scales exponentially with the number of codeword bits.
If on the other hand, a small gap to the exact ML performance is acceptable, then different branch-and-bound algorithms such as the sphere decoder (SD) have been known to accept reduced computational resources.
Despite the reduced complexity of sphere decoding, recent work in \cite{JE:11} has revealed that, to achieve a vanishing error-gap to optimal ML solutions, even such branch-and-bound algorithms generally require computational resources that, albeit significantly smaller than those required by a brute-force ML decoder, again grow exponentially in the rate and the dimensionality, and remain prohibitive for several MIMO scenarios.

This high complexity required by ML-based decoding solutions, serves as further motivation for exploring other families of decoding methods.  A natural alternative is lattice decoding obtained by simply removing the constellation boundaries of the ML-based search, an action that loosely speaking exploits a certain symmetry which in turn may yield faster implementations. It is the case though that even with lattice decoding, the computational complexity can be prohibitive: finding the exact solution to the lattice decoding problem is generally an NP hard problem (cf. \cite{DM:01}).  At the same time though, the other extreme of very early terminations of lattice decoding, such as linear solutions, have been known to achieve computational efficiency at the expense though of a very sizable, and often unbounded, gap to the exact solution of the lattice decoding problem.

In this work we explore lattice decoding solutions that, in conjunction with terminating policies, strike the proper balance between this exponential complexity and exponential gap.
\subsection{System model}
We consider the general $m \times n$ point-to-point multiple-input multiple-output model given by
\begin{align}
\mathbf{y}=\sqrt{\rho}\mathbf{H}\mathbf{x}+\mathbf{w}
\label{eq:sys}
\end{align}
where $\mathbf{x} \in \mathbb{R}^{m}$, $\mathbf{y} \in \mathbb{R}^{n}$ and $\mathbf{w} \in \mathbb{R}^{n}$ respectively denote the transmitted codewords, the received signal vectors, and the additive white Gaussian noise with unit variance, where the parameter $\rho$ takes the role of the signal to noise ratio (SNR), and where the fading matrix $\mathbf{H} \in \mathbb{R}^{n\times m}$ is assumed to be random, with elements drawn from arbitrary statistical distributions. We consider that one use of \eqref{eq:sys} corresponds to $T$ uses of some underlying ``physical'' channel.  We further assume the transmitted codewords $\mathbf{x}$ to be uniformly distributed over some codebook $\mathcal{X} \in \mathbb{R}^{m}$, to be statistically independent of the channel $\mathbf{H}$, and to satisfy the power constraint
\begin{align}
E\{\left\|\mathbf{x}\right\|^2\} \le T.
\label{eq:power}
\end{align}

\subsection{Rate, reliability and complexity in outage-limited MIMO communications}
In terms of error performance, we let $P_e$ denote the probability of codeword error, and we consider the rate,
\begin{align}
R=\frac{1}{T} \log |\mathcal{X}|,
\label{eq:rate}
\end{align}
in bits per channel use (bpcu), where $|\mathcal{X}|$ denotes the cardinality of $\mathcal{X}$.

Regarding complexity, we let $N_{\max}$ describe the computational resources, in floating point operations (flops) per $T$ channel uses, that the transceiver is endowed with, in the sense that after $N_{\max}$ flops, the transceiver must simply terminate, potentially prematurely and before completion of its task.
We note that naturally, $N_{\max}$ is intimately intertwined with the desired $P_e$ and $R$, and that any attempt to significantly reduce $N_{\max}$ may be at the expense of a substantial degradation in error-performance.

In the high SNR regime, a given encoder $\mathcal{X}_r$ and decoder $\mathcal{D}_r$ are said to achieve a \emph{multiplexing gain} $r$ (cf. \cite{ZT:03}) and \emph{diversity gain} $d(r)$ if
\begin{align} \label{eq:DMT1}\displaystyle\lim_{\rho\to\infty}\frac{R(\rho)}{\log \rho} = r, \qquad \text{and} \qquad -\displaystyle\lim_{\rho\to\infty}\frac{\log P_e}{\log \rho} = d(r).\end{align}

In the same high SNR regime, the complexity is here chosen to take the form
\begin{align} \label{eq:complexityExponent}
c(r) : =   \lim_{\rho \rightarrow \infty} \frac{N_{\max}}{\log\rho},
\end{align} which is henceforth denoted as the \emph{complexity exponent}.
Noting that $R = r\log\rho$, we observe that $c(r)>0$ implies a complexity that is exponential in the rate.
\begin{remark}
A reasonable question at this point would pertain as to why the computational resources $N_{\max}$ scale with $\rho$ and are dependent on $r$, to which we note that the complexity of decoding is generally dependent on the density of the codebook, which in turn depends on $\rho$ and $R$. Furthermore this dependence of the complexity exponent (and by extension of $N_{\max}$) on $r$, reflects a potential ability to regulate the computational resources depending on the rate.  Finally the fact that both $P_e$ and $N_{\max}$ are represented as polynomial functions of $\rho$, simply stems from the fact that both $P_e$ and $|\mathcal{X}|$ naturally scale as polynomial functions of $\rho$.  Specifically we quickly note that $c(r)$ captures the entire complexity range
\[0\leq c(r)\leq rT \]
of all reasonable transceivers, with $c(r)=0$ corresponding to the fastest possible transceiver (requiring a subexponential number of flops per $T$ channel uses), and with $c(r)= rT$ corresponding to the optimal but arguably slowest, full-search uninterrupted ML decoder\footnote{We here note that strictly speaking, $\mathcal{X}_r,\mathcal{D}_r$ may potentially introduce a complexity exponent larger than $rT$.  In such a case though, $\mathcal{X}_r,\mathcal{D}_r$ may be substituted by a lookup table implementation of $\mathcal{X}_r$ and an unrestricted ML decoder.  This encoder-decoder will jointly require resources that are a constant multiple of $|\mathcal{X}_r|\doteq \rho^{rT}$ as it has to construct and visit all possible $|\mathcal{X}_r|$ codewords, at a computational cost of a bounded number of flops per codeword visit.  It is noted that the number of flops per visited codeword is naturally independent of $\rho$.} in the presence of a canonical code with multiplexing gain $r$, i.e., with $|\mathcal{X}_r|  = 2^{RT}= \rho^{rT}$.
\end{remark}
\vspace{5pt}

If this canonical code though is linear, searching the entire codebook can be avoided by algorithmic solutions like the sphere decoder (SD) which can provide substantial complexity reductions at a potential small loss in error performance.  Such solutions take advantage of the linear nature of the code that is defined by a \emph{generator matrix} $\bfG$ and a \emph{shaping region} $\mathcal{R}^{'}$. Specifically for $r \geq 0$, a (sequence of) full-rate linear (lattice) code(s) $\mathcal{X}_r$ is given by $\mathcal{X}_r = \Lambda_r \cap \mathcal{R}^{'}$ where $\Lambda_r \defeq \rho^{\frac{-rT}{\kappa}}\Lambda$ and $\Lambda \defeq \{\mathbf{G}\mathbf{s} \ | \ \mathbf{s} \in \mathbb{Z}^{\kappa}\}$, where $\mathbb{Z}^{\kappa}$ denotes the ${\kappa = \min\{m,n\}}$ dimensional integer lattice, where $\mathcal{R}^{'}$ is a compact convex subset of $\mathbb{R}^{\kappa}$ that is independent of $\rho$, and where $\mathbf{G} \in \mathbb{R}^{m \times \kappa}$ is full rank and independent of $\rho$.
For the class of lattice codes considered here, the codewords take the form
\begin{align}
\mathbf{x}=\rho^{\frac{-rT}{\kappa}}\mathbf{G}\mathbf{s}, \ \ \ \ \mathbf{s} \in \mathbb{S}_{r}^{\kappa}\defeq \mathbb{Z}^{\kappa} \cap \rho^{\frac{rT}{\kappa}}\mathcal{R} ,
\label{eq:code}
\end{align}
where $\mathcal{R} \subset \mathbb{R}^{\kappa}$ is a natural bijection of the shaping region $\mathcal{R}^{'}$ that preserves the code, and where $\mathcal{R}$ contains the all zero vector $\mathbf{0}$.

As noted before, despite the reduced complexity of sphere decoding of such lattice codes (as compared to brute-force ML decoding), recent work in \cite{JE:11} has revealed that even such branch-and-bound algorithms generally require computational resources that grow exponentially in the number of codeword bits and the dimensionality. As an indicative example of this high complexity, we note that the work in~\cite{JE:11} showed that such SD algorithms, when applied for decoding a large family of high-performing codes including all known full-rate DMT optimal codes, over the $\nt\times \nr$ quasi-static MIMO channel with Rayleigh fading and $\nr \geq \nt$, introduce a complexity exponent\footnote{Although premature at this point, we hasten to note for the expert reader that this complexity indeed holds irrespective of the radius updating policy, irrespective of the decoding ordering, and as we will see later on, holds even in the presence of MMSE preprocessing.} of the form
\begin{align}
c(r)=\frac{T}{\nt}\bigl(r(\nt-\left\lfloor{r}\right\rfloor-1) +(\nt\left\lfloor{r}\right\rfloor -r(\nt-1))^+\bigr).
\label{eq:cr_ML}
\end{align}
In the above, $\left\lfloor{r}\right\rfloor$ denotes the largest integer not greater than $r$.
The exponent, which simplifies to $c(r)=\frac{T}{\nt}r(\nt-r)$ for integer values of $r$, reaches at $r = \nt/2$ (for even values of $\nt$) an overall maximum value of $\nt T/4$ which, for the aforementioned codes is equal to $\kappa/8$,
corresponding to complexity in the order of
$2^{\frac{1}{8} \kappa \log\rho } = \rho^{\kappa/8} = \sqrt{|\mathcal{X}|}$.  At any fixed multiplexing gain, these required computational resources can be seen to be in the order of $2^{RT(\frac{\nt - r}{\nt})}$ flops which reveals a complexity that is exponential in the number of codeword bits, and a corresponding exponential slope of $\frac{\nt - r}{\nt}$.

\subsection{Transition to lattice decoding for reducing complexity}
As mentioned, this high complexity of ML based (constrained) decoders, motivates consideration of other decoder families, with a natural alternative being the unconstrained (naive) lattice decoder which takes the general form
\begin{align}
\mathbf{\hat{x}}_{L}=\arg \min_{\mathbf{\hat{x}} \in \Lambda_{r}} \left\|\mathbf{y}-\sqrt{\rho}\mathbf{H}\mathbf{\hat{x}}\right\|^2.
\label{eq:lattice}
\end{align}
Naturally when $\mathbf{\hat{x}}_{L} \notin \mathcal{X}_{r}$, the decoder declares an error.

The use of lattice decoding, and specifically of preprocessed lattice decoding in MIMO communications has received substantial attention from works like \cite{AEV+:02}, \cite{MGD+:06} and \cite{GCD:04}, where the latter proved that lattice decoding in the presence of MMSE preprocessing achieves the optimal DMT for specific MIMO channels and statistics, and for DMT-optimal random codes.  The use of lattice decoding as an alternative to computationally expensive ML based solutions, was recently further validated on the one hand by the aforementioned work in \cite{JE:11}, \cite{JE:11isit} which revealed the large computational disadvantages of ML based solutions, and on the other hand by the work in~\cite{JE:10} which further confirmed the performance advantages of lattice decoding by showing that regularized (MMSE-preprocessed)\footnote{We will interchangeably use \emph{MMSE-preprocessed decoder} and \emph{regularized decoder}, with the first term being more commonly used, and with the second implying a more general family of decoders (cf.~\cite{JE:10} where the equivalence between the two decoders is discussed.).
Even though in the asymptotic setting of interest, the two accept the same results throughout the paper, some extra error-performance gains can be achieved by proper optimization of the regularized decoder (cf.~\cite{PM:11}).} lattice decoding achieves the optimal DMT performance, for almost all MIMO scenarios and fading statistics, and all non-random lattice codes, irrespective of the codes' ML performance.

It is the case though that the aforementioned extreme complexity of exact lattice decoding solutions, in conjunction with the potentially unbounded error-performance degradation (gap) of very early terminations (as opposed to exact implementations) of lattice decoding, bring to the fore the need for balanced approximations of lattice decoding solutions that better balance the very sizable complexity and gap.  Specifically for any simplified variant $\mathcal{D}_r$ of the baseline (exact) MMSE-preprocessed lattice decoder, this gap can, in the high SNR regime, be quantified as
\begin{align}\label{eq:gap1}
g_{L}(c) \defeq \displaystyle\lim_{\rho\to\infty}\frac{P_e}{\prob{\mathbf{\hat{x}}\neq \mathbf{x}}}
\end{align}
where $\prob{\mathbf{\hat{x}}\neq \mathbf{x}}$ describes the probability of error of the \emph{exact} MMSE-preprocessed lattice decoder, where $P_e$ denotes the probability of error of $\mathcal{D}_r$, and where $c$  \ (i.e., $c(r)$) is the complexity exponent that describes the (asymptotic rate of increase of the) computational resources required to achieve this performance gap.  Generally a smaller computational complexity exponent $c$ implies a larger gap $g_{L}(c)$.
The clear task has remained for some time to construct decoders that optimally traverse this tradeoff between $g$ and $c$, i.e., that reduce the performance gap to the exact lattice decoding solution, with reasonable computational complexity.  Equivalently for $N_{\max}(g)$ denoting the computational resources in flops required to achieve a certain gap $g$ to the baseline exact MMSE-preprocessed lattice decoder, the above task can be described, in the high SNR regime, as trying to minimize
\begin{align*}
\displaystyle\lim_{\rho\to\infty}\frac{\log N_{\max}(g)}{\log\rho}.
\end{align*}
This will be achieved later on.

\subsection{Contributions}
We first show that the computational complexity required by the MMSE-preprocessed (unconstrained) lattice sphere decoder, asymptotically matches the complexity of the (constrained) ML-based (MMSE-preprocessed or not) sphere decoders, and is commonly exponential in the dimensionality and the number of codeword bits.  This is established for a large class of codes of arbitrary error-performance, a large class of fading statistics, and specifically for the quasi-static MIMO channel -- for example the complexity required for DMT optimal lattice sphere decoding, in the presence of a large family of DMT optimal codes, takes the previously seen simple piecewise linear form in \eqref{eq:cr_ML}.  In a parenthetical note, and deviating slightly from the spirit of this paper, we also provide a universal upper bound on the complexity of regularized lattice sphere decoding, which holds irrespective of the lattice code applied and irrespective of the fading statistics.  This upper bound again takes the form in \eqref{eq:cr_ML}, matching that in the case of constrained ML-based sphere decoding, thus revealing the surprising fact that there exists no statistical channel behavior that will allow the removal of the bounding region to cause unbounded increases in the complexity of the decoder\footnote{In other words, this complexity bound holds even if the channel statistics are such that the channel realizations cause the decoder to always have to solve the hardest possible lattice search problem.}.

With provable evidence of the very high complexity of regularized lattice decoding, we turn to the powerful tool of lattice reduction and seek to understand its effects on computational complexity.  While there has existed a general agreement in the community that lattice reduction does reduce complexity, cf.~\cite{DGC:03}, this has not yet been supported analytically in any relevant communication settings.
In fact, and quite opposite to common wisdom, it was recently shown that for a fixed-radius\footnote{The radius here is considered fixed in the sense that it does not vary with respect to the channel realization and rate.} sphere decoding implementation of the naive lattice decoder~\cite{SJS+:09}, LR does not improve the sphere decoder complexity tail exponent.

What our present work shows is that lattice reduction reduces an ML-like exponentially increasing complexity, to very manageable subexponential values.
We specifically proceed to prove that the LR-aided regularized lattice decoder, implemented by a fixed-radius sphere decoder and timeout policies that occasionally abort decoding and declare an error, achieves
\begin{align*}
g_{L}(\epsilon) =1,\ \ \ \displaystyle\lim_{\rho\to\infty}\frac{\log N_{\max}(g)}{\log\rho}  = 0 \ \ \ \ \ \forall \epsilon>0,g\geq 1,
\end{align*}
i.e., achieves a vanishing gap to the exact implementation of regularized lattice decoding and does so with a complexity exponent that vanishes to zero, which in turn implies subexponential complexity in the sense that the complexity scales slower than any conceivable exponential function.
It is finally noted that this vanishing gap approach serves the practical purpose of an analytical refinement over basic diversity analysis which generally fails to address potentially massive gaps between theory and practice.
\subsection{Notation}
We use $\doteq$ to denote the \emph{exponential equality}, i.e., we write $f(\rho)\doteq \rho^{B}$ to denote $\displaystyle\lim_{\rho\to\infty}\frac{\log f(\rho)}{\log \rho}=B$, and $ \dotleq $, $ \dotgeq $ are similarly defined. With this notation, we can write $P_e \doteq \rho^{-d(r)}$ (cf.~\eqref{eq:DMT1}). In this paper we use $\ulcorner\bullet\urcorner$ to denote the smallest integer not smaller than the argument, $\llcorner\bullet\lrcorner$ to denote the largest integer not larger than the argument, $(\bullet)^{H}$ to denote the conjugate transpose of $(\bullet)$, $(\bullet)^{+}$ to denote $\max\{ {0, (\bullet)}\}$ and $vec(\bullet)$ to denote the operation whereby the columns of the argument $(\bullet)$ are stacked to form a vector.

\section{MMSE-Preprocessed Lattice Sphere Decoding Complexity }

We proceed to describe the preprocessed lattice decoder, its sphere decoding implementation, and for a practical setting of interest that includes the quasi-static MIMO channel and common codes, to establish the decoder's computational complexity.
\subsection{Lattice sphere decoding}

Combining \eqref{eq:sys} and \eqref{eq:code} yields the equivalent model
\begin{align}
\mathbf{y}=\mathbf{M}_r\mathbf{s}+\mathbf{w}
\label{eq:newvect}
\end{align}
where
\begin{align}
\mathbf{M}_r= \rho^{\frac{1}{2}-\frac{rT}{\kappa}}\mathbf{H}\mathbf{G} \in \mathbb{R}^{n\times \kappa}
\label{eq:m}
\end{align}
is a function of the multiplexing gain\footnote{For simplicity of notation we will, in most cases, denote $\mathbf{M}_r$ with $\mathbf{M}$.} $r$.

Consequently the corresponding naive lattice decoder in \eqref{eq:lattice} takes the form (see for example~\cite{JE:10}, also \cite{DGC:03})
\begin{align}
\mathbf{\hat{s}}_{L}=\arg \min_{\mathbf{\hat{s}} \in \mathbb{Z}^{\kappa}} \left\|\mathbf{y}-\mathbf{M}\mathbf{\hat{s}}\right\|^2.
\label{eq:nl}
\end{align}
As a result though of neglecting the boundary region, the above decoder declares additional errors if $\mathbf{\hat{s}}_{L} \notin \mathbb{S}_{r}^{\kappa}$, resulting in possible performance costs.  These costs motivated the use of MMSE preprocessing which essentially regularizes the decision metric to penalize vectors outside the boundary constraint $\mathbb{S}_{r}^{\kappa}$ (cf.~\cite{JE:10}).
Specifically the MMSE-preprocessed lattice decoder is obtained by implementing an unconstrained search over the MMSE-preprocessed lattice, and takes the form
\begin{align}
\mathbf{\hat{s}}_{r-ld}=\arg \min_{\mathbf{\hat{s}} \in \mathbb{Z}^{\kappa}} \left\|\mathbf{Fy}-\mathbf{R}\mathbf{\hat{s}}\right\|^2,
\label{eq:mmse_rld}
\end{align}
where $\mathbf{F}$ and $\mathbf{R}$ are respectively the MMSE forward and feedback filters such that $\mathbf{F}=\mathbf{R}^{-H}\mathbf{M}^{H}$, where \begin{align}\label{eq:feedback}\mathbf{R}^H\mathbf{R}=\mathbf{M}^H\mathbf{M}+\alpha_r^2\mathbf{I},\end{align} where $\alpha_r = \rho^{\frac{-rT}{\kappa}}$ and where $\mathbf{R}$ is an upper-triangular matrix (more details can be found in Appendix \ref{app:noise_vec}). For $\mathbf{r} \defeq \mathbf{F}\mathbf{y}$, the model transitions from \eqref{eq:newvect} to
\begin{eqnarray}
\mathbf{r} &=& \mathbf{R}^{-H}\mathbf{M}^{H}\mathbf{M}\mathbf{s}+\mathbf{R}^{-H}\mathbf{M}^{H}\mathbf{w} \nonumber \\
&=& \mathbf{R}^{-H}(\mathbf{R}^{H}\mathbf{R} - \alpha_r^2\mathbf{I})\mathbf{s}+\mathbf{R}^{-H}\mathbf{M}^{H}\mathbf{w} \nonumber \\
&=& \mathbf{R}\mathbf{s}-{\alpha_r}^2\mathbf{R}^{-H}\mathbf{s}+\mathbf{R}^{-H}\mathbf{M}^{H}\mathbf{w} \nonumber \\
&=& \mathbf{R}\mathbf{s}+\mathbf{w'}
\label{eq:ml_mmser}
\end{eqnarray}
where \begin{align}
\label{eq:equivNoise}\mathbf{w}^{'}=-\alpha_r^2\mathbf{R}^{-H}\mathbf{s}+\mathbf{R}^{-H}\mathbf{M}^{H}\mathbf{w}\end{align}
is the equivalent noise that includes self-interference (first summand) and colored Gaussian noise.
Consequently the corresponding regularized lattice decoder takes the form
\begin{align}
\mathbf{\hat{s}}_{r-ld}=\arg \min_{\mathbf{\hat{s}} \in \mathbb{Z}^{\kappa}} \left\|\mathbf{r}-\mathbf{R}\mathbf{\hat{s}}\right\|^2,
\label{eq:rld_mmse}
\end{align}
which is then solved by the sphere decoder which recursively enumerates all lattice vectors $\mathbf{\hat{s}} \in \mathbb{Z}^{\kappa}$ within a given sphere of radius $\xi > 0$, i.e., which identifies as candidates the vectors $\mathbf{\hat{s}}$ that satisfy
\begin{align}
\|\mathbf{r}-\mathbf{R}\mathbf{\hat{s}}\|^2 \leq \xi^2.
\label{eq:sphere1}
\end{align}
The algorithm specifically uses the upper-triangular nature of $\mathbf{R}$ to recursively identify partial symbol vectors $\mathbf{\hat{s}}_k$, $k=1,\cdots,\kappa$, for which
\begin{align}
\|\mathbf{r}_k-\mathbf{R}_{k}\mathbf{\hat{s}}_k\|^2 \leq \xi^2,
\label{eq:sphere2}
\end{align}
where $\mathbf{\hat{s}}_k$ and $\mathbf{r}_k$ respectively denote the last $k$ components of $\mathbf{\hat{s}}$ and $\mathbf{r}$, and where $\mathbf{R}_{k}$ denotes the $k \times k$ lower-right submatrix of $\mathbf{R}$. Clearly any set of vectors $\mathbf{\hat{s}} \in \mathbb{Z}^{\kappa}$, with common last $k$ components that fail to satisfy \eqref{eq:sphere2}, may be excluded from the set of candidate vectors that satisfy~\eqref{eq:sphere1}.

The enumeration of partial symbol vectors $\mathbf{\hat{s}}_k$ is equivalent to the traversal of a regular tree with $\kappa$ layers -- one layer per symbol component of the symbol vectors, such that layer $k$ corresponds to the $k$th component of the transmitted symbol vector\footnote{We will henceforth refer to the symbol vector $\mathbf{s}\in \mathbb{S}_{r}^{\kappa}$ corresponding to the transmitted codeword $\mathbf{x}=\rho^{\frac{-rT}{\kappa}}\mathbf{G}\mathbf{s}$ (cf.~\eqref{eq:code})
, simply as the \emph{transmitted symbol vector}.} $\mathbf{s}$. There is a one-to-one correspondence between the nodes at layer $k$ and the partial vectors $\mathbf{\hat{s}}_k$. We say that a node is visited by the sphere decoder if and only if the corresponding partial vector $\mathbf{\hat{s}}_k$ satisfies \eqref{eq:sphere2}, i.e., there is a bijection between the visited nodes at layer $k$ and the set
\begin{equation}
\mathcal{N}_k  \defeq  \{  \mathbf{\hat{s}}_k \in \mathbb{Z}^k \ | \  \|\mathbf{r}_k-\mathbf{R}_{k}\mathbf{\hat{s}}_k\|^2 \leq \xi^2 \} .
\label{eq:nodes_per_layer}
\end{equation}

\subsection{Complexity of MMSE-preprocessed lattice sphere decoding}
Consequently the total number of visited nodes (in all layers of the tree) is given by
\begin{equation}
N_{SD} = \sum_{k=1}^{\kappa} N_{k}  ,
\label{eq:sphere_complexity}
\end{equation}
where $N_{k} \defeq |\mathcal{N}_k|$ is the number of visited nodes at layer $k$ of the search tree.
The total number of visited nodes is commonly taken as a measure of the sphere decoder complexity.
It is easy to show that in the scale of interest the SD complexity exponent $c(r)$ would not change if instead of considering the number of visited nodes, we considered the number of flops spent by the decoder\footnote{To see this, we consider that the cost of visiting a node, is independent of $\rho$.  Once at a visited node, this same bounded cost includes the cost of establishing which children-nodes not to visit in the next layer.}.

Naturally the total number of visited nodes is a function of the search radius $\xi$.
We here use a fixed radius, which may result in a non-zero probability that the transmitted symbol vector $\mathbf{s}$ is not in $\mathcal{N}_{\kappa}$. Consequently we must choose a radius that strikes the proper balance between decreasing the aforementioned probability and at the same time sufficiently decreasing the size of $\mathcal{N}_{\kappa}$.  Towards this we note that for the transmitted symbol vector $\mathbf{s}$, the metric in~\eqref{eq:rld_mmse} satisfies
\[\|\mathbf{r}-\mathbf{R}\mathbf{s}\|^2 = \| \mathbf{w}^{'}\|^2,\] which means that if $\| \mathbf{w}^{'}\|> \xi$, then the transmitted symbol vector is excluded from the search, resulting in a decoding error.
As Lemma~\ref{lem:noise_vec} will later argue taking into consideration the self-interference and non-Gaussianity of $\mathbf{w}^{'}$, we can set $\xi = \sqrt{z \log \rho}$, for some $z > d(r)$ such that  \[\prob{\|\mathbf{w}^{'}\|^2 > \xi^2} \ \dot < \ \rho^{-d(r)},\] which implies a vanishing probability of excluding the transmitted information vector from the search, and a vanishing degradation of error performance.

We here note that the MMSE-preprocessed lattice sphere decoder differs from its ML-based equivalent in two aspects: the presence of MMSE preprocessing and the absence of a bounding region to constrain the search.  These two aspects are generally perceived to have an opposite effect on the complexity.  On the one hand, MMSE preprocessing, which we recall from~\eqref{eq:nodes_per_layer} to introduce unpruned sets
\begin{equation*}
\mathcal{N}_k  \defeq  \{  \mathbf{\hat{s}}_k \in \mathbb{Z}^k  \  |  \  \|\mathbf{r}_k-\mathbf{R}_{k}\mathbf{\hat{s}}_k\|^2 \leq \xi^2 \}, \ \ k=1,\cdots,\kappa,
\end{equation*}
is associated to reduced complexity in lattice-based SD solutions (cf.~\cite{SJS+:09}) due to the resulting penalization of faraway lattice points (cf.~\cite{JE:10}).  On the other hand, the absence of boundary constraints can be associated to increased complexity as it introduces an unbounded number of candidate vectors.
We proceed to show that in terms of the complexity exponent, under common MIMO scenarios and codes, these two aspects exactly cancel each other out, and that consequently MMSE-preprocessed lattice sphere decoding introduces a complexity exponent that matches that of ML-based sphere decoding (cf. \cite{JE:11}), which it self is shown here to also match the complexity exponent of ML-based SD in the presence of MMSE preprocessing\footnote{We clarify that ML-based SD in the presence of MMSE preprocessing, corresponds to unpruned sets $\mathcal{N}_k \cap \mathbb{S}_{r}^{k}$ where $\mathbb{S}_{r}^{k}$ is the $k$-dimensional set resulting from the natural reduction of $\mathbb{S}_{r}^{\kappa}$ from \eqref{eq:code}.}.

Before proceeding we note that this analysis is specific to sphere decoding, and that it does not account for any other ML based solutions that could, under some (arguably rare) circumstances, be more efficient.  A classical example of such rare circumstances would be a MIMO scenario, or equivalently a set of fade statistics, that always generate diagonal channel matrices.  Another example would be having codes drawn from orthogonal designs which introduce very small decoding complexity, but which are provably shown to be highly suboptimal except for very few unique cases like the $\nt=2,\nr=1$ quasi-static case~\cite{Rad:23}.
In light of this, in this section only, we mainly focus on the widely considered $n_T\times n_R$ ($n_R \geq n_T$) i.i.d. and quasi-static MIMO setting and on the large but specific family of full-rate ($\kappa = 2\min\{\nt,\nr\}T = 2\nt T$) threaded codes (cf. \cite{GD:03,SRS:03,EKP+:06,ORB:06}), which includes all known DMT optimal codes as well as uncoded transmission (V-BLAST).

We proceed with the main Theorem of the section, which applies under natural detection ordering (cf.~\cite{MGD+:06,JE:11}), and under the assumption of i.i.d. regular fading statistics\footnote{The i.i.d. regular fading statistics satisfy the general set of conditions as described in \cite{ZMM:07}, where a) the near-zero behavior of the fading coefficients $h$ is bounded in probability as $c_1|h|^t \leq  p(h)  \leq c_2|h|^t$ for some positive and finite $c_1$, $c_2$ and $t$, where b) the tail behavior of $h$ is bounded in probability as $p(h)  \leq c_2e^{-b|h|^\beta}$ for some positive and finite $c_2$, $b$ and $\beta$, and where c) $p(h)$ is upper bounded by a constant $K$.}.

\begin{theorem}
\label{thm:cr_mmse}
The complexity exponent for MMSE-preprocessed lattice sphere decoding any full-rate threaded code over the quasi-static MIMO channel with i.i.d. regular fading statistics, is equal to the complexity exponent of ML-based SD with or without MMSE preprocessing.
\end{theorem}

\vspace{3pt}
\begin{IEEEproof}
See Appendix~\ref{app:cr_mmse}.
\end{IEEEproof}
\vspace{3pt}

We clarify that even though all three decoders are DMT optimal, the above result incorporates more than just DMT optimal decoding, in the sense that any timeout policy will tradeoff $d(r)$ with $c(r)$ identically for ML-based and lattice-based sphere decoding.  In other words the three decoders share the same $d(r)$ and $c(r)$ capabilities, irrespective of the timeout policy.

Furthermore, considering different SD detection orderings (cf. \cite{MGD+:06}), the following extends the range of codes for which the ML-based and lattice-based SD share a similar complexity.  The proof follows from the proof of Theorem~\ref{thm:cr_mmse} in Appendix~\ref{app:cr_mmse}, and from Theorem 4 in \cite{JE:11}.

\begin{corollary}
\label{thm:cr_mmse_ordering}
Given any full-rate code of arbitrary DMT performance,  there is always at least one non-random fixed permutation of the columns of $\bfG$, for which the complexity exponent of the MMSE-preprocessed lattice sphere decoder matches that of the ML based sphere decoder.
\end{corollary}
\vspace{3pt}

The following focuses on a specific example of practical interest.
\begin{corollary}
\label{thm:cr_mmse1}
The complexity exponent for DMT optimal MMSE-preprocessed lattice sphere decoding of minimum delay $(T=n_T)$ DMT optimal threaded codes over the quasi-static MIMO channel with i.i.d. regular fading statistics, takes the following form
\begin{align}
c_{r-ld}(r)=r(n_T-\left\lfloor{r}\right\rfloor-1) +(n_T\left\lfloor{r}\right\rfloor -r(n_T-1))^+,
\label{eq:cr_mmse}
\end{align}
which simplifies to
\begin{align}
c_{r-ld}(r)=r(n_T-r)
\label{eq:cr_mmseSimpleMinDelay}
\end{align}
for integer values of $r$.
\end{corollary}
\vspace{3pt}
\begin{IEEEproof}
See Appendix~\ref{app:corollaries}.
\end{IEEEproof}
\vspace{3pt}

Further evidence that connects the complexity behavior of MMSE-preprocessed lattice-based SD, with that of its ML-based counterpart, now comes in the form of a non-trivial universal bound that is shared by the two methods. This is particularly relevant because unconstrained lattice decoding could conceivably require unbounded computational resources given the unbounded number of candidate lattice points.  Specifically the following universal upper bound on the complexity of regularized lattice-based SD, matches the upper bound in \cite{JE:11} for the ML case, and it holds irrespective of the full-rate lattice code applied and irrespective of the fading statistics. The generality with respect to the fading statistics is important because it guarantees that no set of fading statistics, even those that always generate infinitely dense lattices, can cause an unbounded increase in the complexity due to removal of the boundary constraints.

\begin{corollary}
\label{thm:cr_mmse_bound}
Irrespective of the fading statistics and of the full-rate lattice code applied, the complexity exponents of MMSE-preprocessed lattice SD and of ML-based SD, are upper bounded by
\begin{align}
\overline{c}(r)=\frac{T}{n_T}\left(r(n_T-\left\lfloor{r}\right\rfloor-1) +(n_T\left\lfloor{r}\right\rfloor -r(n_T-1))^+\right)
\label{eq:bound_cr_mmse}
\end{align}
which simplifies to
\begin{align}
\overline{c}(r)=\frac{T}{n_T}r(n_T-r)
\end{align}
for integer $r$.
\end{corollary}
\vspace{3pt}

%%%%%%%%%%%%%%%%%%%%%%%%%%%%%%%%%%%%%%%%%%%%%%%%%%%%%%%%%%%%%%%%%%%%%%%%%%%%%%%%%%%%%%%%%%%%
\begin{IEEEproof}
See Appendix \ref{app:corollaries}.
\end{IEEEproof}
\vspace{3pt}

The above results revealed the very high, ML-like complexity of MMSE-preprocessed lattice decoding.  Coming back to the main focus of the paper, and after reverting to the most general setting of MIMO scenarios, statistics and full-rate lattice codes, we proceed to show how proper utilization of lattice sphere decoding and LR techniques can indeed reduce the complexity exponent to zero, at an error-performance cost that vanishes in the high SNR limit.

%%%%%%%%%%%%%%%%%%%%%%%%%%%%%%%%%%%%%%%%%%%%%%%%%%%%%%%%%%%%%%%%%%%%%%%%%%%%%%%%%%%%%%%%%%%%%%%%%%%%%%%%%%%%
%%%%%%%%%%%%%%%%%%%%%%%%%%%%%%%%%%%%%%%%%%%%%%%%%%%%%%%%%%%%%%%%%%%%%%%%%%%%%%%%%%%%%%%%%%%%%%%%%%%%%%%%%%%%

\section{LR-aided Regularized Lattice Sphere Decoding Complexity}

Lattice reduction techniques have been typically used in the MIMO setting to improve the error performance of suboptimal decoders (cf. \cite{YW:02}, \cite{WF:03}, see also \cite{WSJ+:11}, \cite{Lin+:11}). In the current setting the LR algorithm, which is employed at the receiver after the action of MMSE preprocessing, modifies the search of the MMSE-preprocessed lattice decoder,
from
\begin{align*}
\mathbf{\hat{s}}_{rld}=\arg \min_{\mathbf{\hat{s}} \in \mathbb{Z}^{\kappa}} \left\|\mathbf{r}-\mathbf{R}\mathbf{\hat{s}}\right\|^2
\end{align*}
(cf. \eqref{eq:rld_mmse}), to the new
\begin{align}
\mathbf{\tilde{s}}_{lr-rld}=\arg \min_{\mathbf{\hat{s}} \in \mathbb{Z}^{\kappa}} \left\|\mathbf{r}-\mathbf{R}\mathbf{T}\mathbf{\hat{s}}\right\|^2,
\label{eq:rld_mmse_lr2}
\end{align}
by accepting as input the MMSE-preprocessed lattice generator matrix $\mathbf{R}$, and producing as output the matrix $\mathbf{T}\in \mathbb{Z}^{\kappa \times \kappa}$ which is unimodular meaning that it has integer coefficients and unit-norm determinant, and which is designed so that $\mathbf{R}\mathbf{T}$ is (loosely speaking) more orthogonal than $\mathbf{R}$. As a result of this unimodularity, we have that $\mathbf{T}^{-1}\mathbb{Z}^{\kappa} = \mathbb{Z}^{\kappa}$, and consequently the new search in \eqref{eq:rld_mmse_lr2}
corresponds to yet another \emph{lattice} decoder, referred to as the LR-aided MMSE-preprocessed lattice decoder,
which operates over a generally better conditioned channel matrix $\mathbf{R}\mathbf{T}$.

Finally with sphere decoding in mind, the LR algorithm is followed by the QR decomposition\footnote{A more proper statement would be that the QR decomposition is performed by the LR algorithm it self.} of the new lattice-reduced MMSE-preprocessed matrix $\mathbf{R}\mathbf{T}$, resulting in a new upper-triangular model
\begin{eqnarray}
\mathbf{\tilde{r}} &=& \mathbf{\tilde{R}}\mathbf{\tilde{s}}+\mathbf{w''}
\label{eq:lr_ml_mmser}
\end{eqnarray}
and in the new LR-aided MMSE-preprocessed lattice search, which accepts the application of the sphere decoder, and which takes the form
\begin{align}
\mathbf{\tilde{s}}_{lr-rld}=\arg \min_{\mathbf{\hat{s}} \in \mathbb{Z}^{\kappa}} \left\|\mathbf{\tilde{r}}-\mathbf{\tilde{R}}\mathbf{\hat{s}}\right\|^2,
\label{eq:rld_mmse_lr1}
\end{align}
where $\mathbf{\tilde{Q}}\mathbf{\tilde{R}}=\mathbf{R}\mathbf{T}$ corresponds to the QR-decomposition of $\mathbf{R}\mathbf{T}$, where $\mathbf{\tilde{R}}$ is upper-triangular, where $\mathbf{\tilde{r}} \defeq \mathbf{\tilde{Q}}^H\mathbf{r}$,
$\mathbf{\tilde{s}}=\mathbf{T}^{-1}\mathbf{s}$, and where $\mathbf{w''}=\mathbf{\tilde{Q}}^H \mathbf{w'}$.

At the very end,
\begin{align}
\mathbf{\hat{s}}_{lr-rld}=\mathbf{T}\mathbf{\tilde{s}}_{lr-rld},
\label{eq:rld_mmse_lr}
\end{align}
allows for calculation of the estimate of the transmitted symbol vector $\mathbf{s}$ in \eqref{eq:newvect}.

We note here that this (exact) solution of the LR-aided MMSE-preprocessed lattice decoder defined by \eqref{eq:rld_mmse_lr1}, \eqref{eq:rld_mmse_lr}, is identical to the exact solution of the MMSE-preprocessed lattice decoder given by \eqref{eq:rld_mmse}, because
\begin{eqnarray}
\min_{\mathbf{\hat{s}} \in \mathbb{Z}^{\kappa}} \left\|\mathbf{r}-\mathbf{R}\mathbf{\hat{s}}\right\|^2 &=& \min_{\mathbf{\hat{s}} \in \mathbb{Z}^{\kappa}} \left\|\mathbf{r}-\mathbf{R}\mathbf{T}\mathbf{T}^{-1}\mathbf{\hat{s}}\right\|^2 \nonumber \\
&\stackrel{(a)}{=}& \min_{\mathbf{\hat{s}} \in \mathbb{Z}^{\kappa}} \left\|\mathbf{r}-\mathbf{\tilde{Q}}\mathbf{\tilde{R}}\mathbf{T}^{-1}\mathbf{\hat{s}}\right\|^2 \nonumber \\
&\stackrel{(b)}{=}& \min_{\mathbf{\hat{s}} \in \mathbb{Z}^{\kappa}} \left\|\mathbf{\tilde{r}}-\mathbf{\tilde{R}}\mathbf{T}^{-1}\mathbf{\hat{s}}\right\|^2 \nonumber \\
&=& \min_{\mathbf{\hat{s}} \in \mathbf{T}^{-1}\mathbb{Z}^{\kappa}} \left\|\mathbf{\tilde{r}}-\mathbf{\tilde{R}}\mathbf{\hat{s}}\right\|^2 \nonumber \\
&\stackrel{(c)}{=}& \min_{\mathbf{\hat{s}} \in \mathbb{Z}^{\kappa}} \left\|\mathbf{\tilde{r}}-\mathbf{\tilde{R}}\mathbf{\hat{s}}\right\|^2,
\label{eq:lr-equi-mmse}
\end{eqnarray}
where $(a)$ follows from the fact that $\mathbf{\tilde{Q}}\mathbf{\tilde{R}}=\mathbf{R}\mathbf{T}$, $(b)$ follows from the rotational invariance of the Euclidean norm, and $(c)$ follows from the fact that $\mathbf{T}^{-1}\mathbb{Z}^{\kappa} = \mathbb{Z}^{\kappa}$.

While though the two lattice decoding solutions (with and without LR) provide identical error performance in the setting of exact implementations, we proceed to show that, in terms of complexity, lattice reduction techniques, and specifically a proper utilization of the LLL algorithm~\cite{LLL:82}, can provide dramatic improvements.

%%%%%%%%%%%%%%%%%%%%%%%%%%%%%%%%%%%%%%%%%%%%%%%%%%%%%%%%%%%%%%%%%%%%%%%%%%%%%%%%%%%%%%%%%%%%%%%%%%%%%%
%%%%%%%%%%%%%%%%%%%%%%%%%%%%%%%%%%%%%%%%%%%%%%%%%%%%%%%%%%%%%%%

\subsection{Complexity of the LR-aided regularized lattice sphere decoder}
%\label{sec:complexity}

We are here interested in establishing the complexity of the LR-aided regularized lattice sphere decoder.  Given that the costs of implementing MMSE preprocessing and of implementing the linear transformation in \eqref{eq:rld_mmse_lr} are negligible in the scale of interest\footnote{Even though the work here focuses on decoding, we can also quickly state the obvious fact that the cost of constructing the codewords is also negligible in the scale of interest because it again only involves a finite-dimensional linear transformation (cf.~\eqref{eq:code}).}, we limit our focus on establishing the cost of lattice reduction, and then the cost of the SD implementation of the search in~\eqref{eq:rld_mmse_lr1}.
Starting with the SD complexity, as in~\eqref{eq:nodes_per_layer}, we identify the corresponding unpruned set at layer $k$ to be
\begin{equation}\label{eq:unprunedSetLR}
\mathcal{N}_k  \defeq  \{  \mathbf{\hat{s}}_k \in \mathbb{Z}^k   \ | \    \|\mathbf{\tilde{r}}_k-\mathbf{\tilde{R}}_{k}\mathbf{\hat{s}}_k\|^2 \leq \xi^2 \},
\end{equation}
and in bounding the size of the above, we first focus on understanding the statistical behavior of the $k\times k$ lower-right submatrices $\mathbf{\tilde{R}}_{k}$ of matrix $\mathbf{\tilde{R}}$ ($k=1,\cdots,\kappa$), where we recall that $\mathbf{\tilde{R}}$ is the upper triangular code-channel matrix, after MMSE preprocessing and LLL lattice reduction.
Towards this, and for $d_L(r-\epsilon)$ denoting the diversity gain of the exact implementation of the regularized lattice decoder at multiplexing gain $r-\epsilon$, we have the following lemma on the smallest singular value of $\mathbf{\tilde{R}}_{k}$.  The proof appears in Appendix~\ref{app:lambdamin}.
\begin{lemma}
\label{lem:lambdamin}
The smallest singular value $\sigma_{min}(\mathbf{\tilde{R}}_{k})$ of submatrix $\mathbf{\tilde{R}}_{k}, \ k=1,\cdots,\kappa$, satisfies
\begin{align}\prob{\sigma_{min}(\mathbf{\tilde{R}}_{k}) \stackrel{.}{<} \rho^{\frac{-\epsilon T}{\kappa}}} \dotleq \rho^{-d_{L}(r-\epsilon)} , \ \text{for all}  \ r \geq  \epsilon > 0.\end{align}
\end{lemma}
\vspace{3pt}

To bound the cardinality $N_k$ of $\mathcal{N}_k$ (cf.~\eqref{eq:unprunedSetLR}), and eventually the total number $N_{SD} = \sum_{k=1}^{\kappa} N_k$ of lattice points visited by the SD, we proceed along the lines of the work in \cite{JE:11}, making the proper modifications to account for MMSE preprocessing, for the removal of the bounding region, and for lattice reduction.

Towards this we see that, after removing the boundary constraint, Lemma 1 in \cite{JE:11} tells us that
\begin{align*}
N_k \defeq \left|\mathcal{N}_k\right| \leq \prod_{i=1}^{k}\left[\sqrt{k}+\frac{2\xi}{\sigma_i(\mathbf{\tilde{R}}_{k})}\right],
\end{align*}
where \[ \sigma_{min}(\mathbf{\tilde{R}}_{k}) = \sigma_1(\mathbf{\tilde{R}}_{k})\leq \cdots \leq \sigma_k(\mathbf{\tilde{R}}_{k})\] are the singular values of $\mathbf{\tilde{R}}_{k}$.
Consequently we have that
\begin{eqnarray}
N_k &{\leq}& \left[\sqrt{k}+\frac{2\xi}{\sigma_{min}(\mathbf{\tilde{R}}_{k})}\right]^{k}. \nonumber \\
\label{eq:nk01}
\end{eqnarray}
As a result, for any $\mathbf{\tilde{R}}_k$ such that \begin{align}\label{eq:assumption1}\sigma_{min}(\mathbf{\tilde{R}}_k) \dotgeq \rho^{\frac{-\epsilon T}{\kappa}},\end{align} and given that $\xi = \sqrt{z\log\rho}$ for some finite $z$, then
\begin{align}
N_k{\dotleq} \left(\sqrt{k}+\frac{2\sqrt{z\log\rho}}{\rho^{\frac{-\epsilon T}{\kappa}}}\right)^{k} \doteq \rho^{\frac{\epsilon T k}{\kappa}},
\label{eq:nk1}
\end{align}
which guarantees that the total number of visited lattice points is upper bounded as
\begin{align}
N_{SD} = \sum_{k=1}^{\kappa} N_k \dotleq \sum_{k=1}^{\kappa} \rho^{\frac{\epsilon T k}{\kappa}} \doteq \rho^{\epsilon T}.
\label{eq:n_sd}
\end{align}
Consequently, directly from Lemma~\ref{lem:lambdamin}, we have that
\begin{align} \label{eq:timeoutProbSD}
\prob{ N_{SD} \ \dot \geq \ \rho^{\epsilon T}} \ \dot \leq \ \rho^{-d_L(r-\epsilon)}.
\end{align}
A similar approach deals with the complexity of the LLL algorithm, which is known (cf. \cite{JSM:08}) to be generally unbounded.  Specifically drawing
from~\cite[Lemma 2]{JE:10}, under the natural assumption of power-limited channels\footnote{This is a moderate assumption that asks that $\expt{\| \bfH \|^2\fro} \dotleq \rho$.  We note that this holds true for any telecommunications setting.}
(cf.\cite{JE:10}), under the natural assumption that $d_L(r-\epsilon)>d_L(r)$ for all $\epsilon>0$, and for $N_{LR} $ denoting the number of flops spent by the LLL algorithm, one can readily conclude that
\begin{align}\label{eq:timeoutProbLR}
\prob{ N_{LR} \geq \gamma\log\rho} \ \dot\leq \ \rho^{-d_L(r-\epsilon)},
\end{align}
for any $\gamma>\frac{1}{2}(d_L(r-\epsilon))$.
Consequently the overall complexity $$N\doteq N_{SD}+N_{LR},$$  in flops, for the LR-aided MMSE preprocessed lattice sphere decoder, satisfies the following
\begin{eqnarray}
\prob{N \dot\geq \rho^{\epsilon T}} &\doteq & \prob{\{N_{SD} \dot\geq \rho^{\epsilon T}\} \cup \{N_{LR} \dot\geq \rho^{\epsilon T}\}} \nonumber \\
&\dotleq& \rho^{-d_{L}(r-\epsilon)}.
\label{eq:prob_run_n}
\end{eqnarray}
Now going back to \eqref{eq:complexityExponent}, and having in mind appropriate timeout policies that bound $N_{\max}$ while at the same time specifically guarantee a \emph{vanishing error performance gap} to the exact solution of regularized lattice decoding, we can see that the complexity exponent $c(r)$ takes the equivalent form recently introduced (for the ML case) in \cite{JE:11}
\begin{align}
c(r) =  \inf \{  x  \ | \  - \lim_{\rho \rightarrow \infty} \frac{\log \prob{N \geq \rho^{x}}}{\log \rho} > d_L(r) \}.
\label{eq:complexity_exp2}
\end{align}
To see this we quickly note that for $N_{\max} = \rho^x$ where $x=c(r)-\delta$ for any $\delta>0$, it is the case that (cf.~\eqref{eq:gap1}) $\lim_{\rho \rightarrow \infty} \frac{\prob{N \geq \rho^{x}}}{\prob{\mathbf{\hat{x}}_{L}\neq \mathbf{x}}}\rightarrow \infty$.

Finally applying~\eqref{eq:prob_run_n} we see that for any positive $\epsilon_1<\epsilon$, it is the case that
\begin{align}
c(r) =  \inf \{  \epsilon   \ | \  - \lim_{\rho \rightarrow \infty} \frac{\log \prob{N \geq \rho^{\epsilon T + \epsilon_1 }}}{\log \rho} > d_L(r) \}
\label{eq:complexity_exp3}
\end{align}
which vanishes arbitrarily close to zero, resulting in a zero complexity exponent.

What remains is to consider the error-performance gap in the presence the LR-aided regularized lattice SD with a timeout policy that interrupts at $N_{\max} = \rho^x$ for any vanishingly small $x>0$.

%%%%%%%%%%%%%%%%%%%%%%%%%%%%%%%%%%%%%%%%%%%%%%%%%%%%%%%%%%%%%%%%%%%%%%%%%

\subsection{Gap to the exact solution of MMSE-preprocessed lattice decoding}
\label{sec:gap}
We here prove that the LR-aided regularized lattice sphere decoder and the associated time-out policies that guarantee a vanishing complexity exponent, also guarantee a vanishing gap to the error performance of the exact lattice decoding implementation. This result is motivated by potentially exponential gaps in the performance of other DMT optimal decoders (cf. \cite{JE:10}), where these gaps may grow exponentially up to $2^{\frac{\kappa}{2}}$ (cf. \cite{LB:86}) or may potentially be unbounded~\cite{LL:10}.

Towards establishing this gap, we recall that the exact MMSE-preprocessed lattice decoder in \eqref{eq:mmse_rld} makes errors when $\mathbf{\hat{s}}_{r-ld}\neq \mathbf{s}$. On the other hand the LLL-reduced MMSE-preprocessed lattice sphere decoder with run-time constraints, in addition to making the same errors ($\mathbf{\hat{s}}_{r-lr-ld}\neq \mathbf{s}$), also makes errors when the run-time limit of $\rho^x$ flops becomes active, i.e., when $N\geq \rho^x$, as well as when a small search radius causes $\mathcal{N}_{\kappa}=\emptyset$.  Consequently the corresponding performance gap to the exact regularized decoder, takes the form
\begin{align*}
g_L(x) = \displaystyle\lim_{\rho\to\infty}\frac{\prob{\{\mathbf{\hat{s}}_{r-lr-ld}\neq \mathbf{s}\}\cup \{N \geq \rho^{x}\}\cup\{\mathcal{N}_{\kappa}=\emptyset\}}}{\prob{\mathbf{\hat{s}}_{r-ld}\neq \mathbf{s}}}.
\end{align*}
To bound the above gap, we apply the union bound and the fact that
\[\prob{\mathcal{N}_{\kappa}=\emptyset}\leq \prob{\|\mathbf{w''}\| > \xi}\]
to get that
\begin{eqnarray} \label{eq:unionBoundGap}
g_L(x) &\leq& \displaystyle\lim_{\rho\to\infty}\frac{\prob{\mathbf{\hat{s}}_{r-lr-ld}\neq \mathbf{s}}}{\prob{\mathbf{\hat{s}}_{r-ld}\neq \mathbf{s}}}+\displaystyle\lim_{\rho\to\infty}\frac{\prob{N \geq \rho^{x}}}{\prob{\mathbf{\hat{s}}_{r-ld}\neq \mathbf{s}}} \nonumber \\
&&+ \displaystyle\lim_{\rho\to\infty}\frac{\prob{\|\mathbf{w''}\| > \xi}}{\prob{\mathbf{\hat{s}}_{r-ld}\neq \mathbf{s}}}.
\end{eqnarray}
Furthermore from \eqref{eq:lr-equi-mmse} we observe that
\begin{align}\label{eq:firstTerm} \prob{\mathbf{\hat{s}}_{r-lr-ld}\neq \mathbf{s}} = \prob{\mathbf{\hat{s}}_{r-ld}\neq \mathbf{s}},\end{align}
and from \eqref{eq:prob_run_n} we recall that
\[\prob{N \dot\geq \rho^{\epsilon T}} \dotleq \rho^{-d_{L}(r-\epsilon)}\] which implies that
for any $x>0$ it holds that
\begin{align}\label{eq:secondTerm}\displaystyle\lim_{\rho\to\infty}\frac{\prob{N \geq \rho^{x}}}{\prob{\mathbf{\hat{s}}_{r-ld}\neq \mathbf{s}}} = 0.\end{align}
Finally the last term in \eqref{eq:unionBoundGap} relates to the search radius $\xi$, and to the behavior of the noise $\mathbf{w}^{''}$ which was shown in \eqref{eq:equivNoise}, \eqref{eq:lr_ml_mmser}
to take the form
\begin{align}\label{eq:NoiseEquiv}\mathbf{w''}=\mathbf{\tilde{Q}}^H \left( -\alpha_r^2\mathbf{R}^{-H}\mathbf{s}+\mathbf{R}^{-H}\mathbf{M}^{H}\mathbf{w}\right).\end{align}

The following lemma, whose proof is found in Appendix~\ref{app:noise_vec}, accounts for the fact that $\mathbf{w}^{''}$ includes self-interference and colored noise, to bound the last term in \eqref{eq:unionBoundGap}.

\vspace{3pt}
\begin{lemma}
\label{lem:noise_vec}
There exist a finite $z> d_L(r)$ for which a search radius $\xi = \sqrt{z \log \rho}$ guarantees that
\begin{align}\label{eq:thirdTerm}\displaystyle\lim_{\rho\to\infty}\frac{\prob{\|\mathbf{w''}\| > \xi}}{\prob{\mathbf{\hat{s}}_{r-ld}\neq \mathbf{s}}}=0.\end{align}
\end{lemma}
\vspace{3pt}

Consequently combining \eqref{eq:firstTerm}, \eqref{eq:secondTerm} and \eqref{eq:thirdTerm} gives that $g_L(x) = 1, \ \ \forall x>0$.
The following directly holds.
\vspace{3pt}

\begin{theorem}
\label{thm:gap}
LR-aided MMSE-preprocessed lattice sphere decoding with a computational constraint activated at $\rho^x$ flops, allows for a vanishing gap to the exact solution of MMSE-preprocessed lattice decoding, for any $x>0$. Equivalently the same LR-aided decoder guarantees that
\begin{align*}
g_{L}(\epsilon) =1 \ \ \ \text{and} \ \ \ \ \lim_{\rho\to\infty}\frac{\log N_{\max}(g)}{\log\rho}  = 0 \ \ \ \ \ \forall \epsilon>0,g\geq 1,
\end{align*}
for all fading statistics, all MIMO scenarios, and all full-rate lattice codes.
\end{theorem}

\section{Conclusions}

The work identified the first lattice decoding solution that achieves, in the most general outage-limited MIMO setting and the high rate and high SNR limit, both a vanishing gap to the error-performance of the (DMT optimal) exact solution of preprocessed lattice decoding, as well as a computational complexity that is subexponential in the number of codeword bits.  The proposed solution employs lattice reduction (LR)-aided regularized lattice sphere decoding and proper timeout policies.
As it turns out, lattice reduction is a special ingredient that allows for complexity reductions; a role that was rigorously demonstrated here for the first time, by proving that without lattice reduction, for most common codes, the complexity cost for asymptotically optimal regularized lattice sphere decoding is exponential in the number of codeword bits, and in many cases it in fact matches the complexity cost of ML sphere decoding.

In light of the fact that, prior to this work, a vanishing error performance gap was generally attributed only to near-full lattice searches that have exponential complexity, in conjunction with the fact that subexponential complexity was generally attributed to early-terminated (linear) solutions which have though a performance gap that can be up to exponential in dimension and/or rate, the work constitutes the first proof that subexponential complexity need not come at the cost of exponential reductions in lattice decoding error performance.

%%%%%%%%%%%%%%%%%%%%%%%%%%%%%%%%%%%%%%%%%%%%%%%%%%%%%%%%%%%%%%%%%%%%%%%%%%%%%%%%%%%%%%%%%%%%
%%%%%%%%%%%%%%%%%%%%%%%%%%%%%%% Appendices %%%%%%%%%%%%%%%%%%%%%%%%%%%%%%%%%%%%%%%%%%%%%%%%%

\appendices

\section{Proof for Theorem~\ref{thm:cr_mmse} and Corollary~\ref{thm:cr_mmse_ordering}}
\label{app:cr_mmse}

In the following we begin by providing an upper bound on the complexity exponent of MMSE-preprocessed (unconstrained) lattice sphere decoding, where this bound holds for the general quasi-static MIMO channel, for all fading statistics and for any full-rate lattice code.
We will then proceed to provide a lower bound on the complexity exponent of the same decoder, where this bound, under the extra assumptions of regular i.i.d. fading statistics and of layered codes, will in fact match the above mentioned upper bound to prove the theorem and the associated corollaries.   Before proceeding with the bounds, we describe the $n_T \times n_R$ ($n_R\geq n_T$) quasi-static point-to-point MIMO channel, and its corresponding association to the general MIMO channel model in \eqref{eq:newvect} and metric in \eqref{eq:rld_mmse}.

The aforementioned quasi-static channel model takes the form
\begin{align}
\mathbf{Y}_C=\sqrt{\rho}\mathbf{H}_C\mathbf{X}_C+\mathbf{W}_C,
\label{eq:sys1}
\end{align}
where $\mathbf{X}_C \in \C^{n_T\times T}$, $\mathbf{Y}_C \in \C^{n_R\times T}$ and $\mathbf{W}_C \in \C^{n_R\times T}$ represent the transmitted, received and noise signals over a period of $T$ time slots, and where $\mathbf{H}_C \in \C^{\nr \times \nt}$ represents the matrix of fade coefficients.  The real-valued representation of \eqref{eq:sys1} can be written as
\begin{align}
\mathbf{Y}_R=\sqrt{\rho}\mathbf{H}_R\mathbf{X}_R+\mathbf{W}_R,
\label{eq:sys2}
\end{align}
where $\mathbf{Y}_R= \left[\begin{array}{l l}
\Re{\{\mathbf{Y}_C\}} & -\Im{\{\mathbf{Y}_C\}}\\ \Im{\{\mathbf{Y}_C\}} & \Re{\{\mathbf{Y}_C\}}\\ \end{array}\right] $, $\mathbf{H}_R= \left[\begin{array}{l l}
\Re{\{\mathbf{H}_C\}} & -\Im{\{\mathbf{H}_C\}}\\ \Im{\{\mathbf{H}_C\}} & \Re{\{\mathbf{H}_C\}}\\ \end{array}\right] $, $\mathbf{X}_R= \left[\begin{array}{l l}
\Re{\{\mathbf{X}_C\}} & -\Im{\{\mathbf{X}_C\}}\\ \Im{\{\mathbf{X}_C\}} & \Re{\{\mathbf{X}_C\}}\\ \end{array}\right] $ and $\mathbf{W}_R= \left[\begin{array}{l l}
\Re{\{\mathbf{W}_C\}} & -\Im{\{\mathbf{W}_C\}}\\ \Im{\{\mathbf{W}_C\}} & \Re{\{\mathbf{W}_C\}}\\ \end{array}\right] $, and subsequent vectorization gives the real-valued model
\begin{eqnarray}
\mathbf{y}&=&\sqrt{\rho}(\mathbf{I}_{T} \kron \mathbf{H}_R)\mathbf{x}+\mathbf{w}
%&=&\sqrt{\rho}\mathbf{H}\mathbf{x}+\mathbf{w}
\label{eq:vect1}
\end{eqnarray}
where $\mathbf{y}=vec(\mathbf{Y}_R)$, $\mathbf{x}=vec(\mathbf{X}_R)$, and $\mathbf{w}=vec(\mathbf{W}_R)$. The system model in \eqref{eq:vect1} is of the familiar form
\begin{align}
\mathbf{y}=\sqrt{\rho}\mathbf{H}\mathbf{x}+\mathbf{w}
\label{eq:vect}
\end{align}
as in \eqref{eq:sys} with $m=2n_TT$, $n=2n_RT$, and where
\begin{align}
\mathbf{H}= \mathbf{I}_{T} \kron \mathbf{H}_R.
\label{eq:h_appendix}
\end{align}
As before the vectorized codewords $\mathbf{x}$, associated to the full-rate code, take the form
\begin{align}
\mathbf{x}=\rho^{\frac{-rT}{\kappa}}\mathbf{G}\mathbf{s}, \ \mathbf{s}\in \ints^{\kappa}\cap \rho^{\frac{rT}{\kappa}}\mathcal{R},
\label{eq:code1}
\end{align}
where $\kappa=2\min\{\nt,\nr\}T =2 \nt T = m$, which allows us to rewrite the model as
\begin{align}
\mathbf{y}=\mathbf{M}\mathbf{s}+\mathbf{w},
\label{eq:sys_eq}
\end{align}
for
\begin{align}
\mathbf{M}=\rho^{\frac{1}{2}-\frac{rT}{\kappa}}\mathbf{H}\mathbf{G}=\rho^{\frac{1}{2}-\frac{rT}{\kappa}}(\mathbf{I}_{T} \kron \mathbf{H}_R)\mathbf{G}.
\label{eq:mr_appendix}
\end{align}
Finally the corresponding coherent MMSE-preprocessed lattice decoder for the transmitted symbol vector $\mathbf{s}$, can be expressed to be (cf. ~\eqref{eq:rld_mmse})
\begin{align}
\mathbf{\hat{s}}_{r-ld}=\arg \min_{\mathbf{\hat{s}} \in \mathbb{Z}^{\kappa}} \left\|\mathbf{r}-\mathbf{R}\mathbf{\hat{s}}\right\|^2,
\end{align}
where $\mathbf{r} =\mathbf{Q}_1^H\mathbf{y}$ and $\mathbf{R} \in \mathbb{C}^{\kappa \times \kappa}$ is the upper-triangular matrix, where furthermore both $\mathbf{Q}_1$ and $\mathbf{R}$ result from the thin QR decomposition of the $(n+\kappa) \times \kappa$ dimensional preprocessed channel matrix
\begin{align}
\label{eq:Mreg}
\mathbf{M}^{reg} \defeq \left[\begin{array}{l}
\mathbf{M}\\ \alpha_r\mathbf{I} \\ \end{array}\right] =
\mathbf{Q}\mathbf{R} = \left[\begin{array}{l}
\mathbf{Q}_1\\ \mathbf{Q}_2 \\ \end{array}\right]\mathbf{R}
\end{align}
and where as before $\alpha_r=\rho^{\frac{-rT}{\kappa}}$.

\subsection{Upper bound on complexity of regularized lattice SD\label{app:Upper}}

In establishing the upper bound, we consider Lemma 1 in \cite{JE:11}, which we properly modify to account for MMSE preprocessing and for the removal of the constellation boundaries, and get that the number $N_k$ of nodes visited at layer $k$ by the MMSE-preprocessed lattice sphere decoder, is upper bounded as
\begin{align}
N_k = \left|\mathcal{N}_k\right| \leq \prod_{i=1}^{k}\left[\sqrt{2k}+\frac{2\xi}{\sigma_i(\mathbf{R}_{k})}\right],
\label{eq:nk11}
\end{align}
where $\sigma_i(\mathbf{R}_k)$, $i=1,\cdots,k$ denote the singular values of $\mathbf{R}_k$ in increasing order.

Towards lower bounding $\sigma_i(\mathbf{R}_k)$, we note that
\begin{align}
\sigma_i(\mathbf{R}_{k}) &\geq  \sigma_i(\mathbf{R})= \sigma_i(\mathbf{M}^{reg}) =  \sqrt{\alpha_r^2+\sigma_i(\mathbf{M}^H\mathbf{M})},\label{eq:equ_sing_appendix}\end{align}
where the first inequality makes use of the interlacing property of singular values of sub-matrices \cite{HJ:85}.
Furthermore for
\begin{align}
\mu_j \defeq -\frac{\log \sigma_j(\mathbf{H}_C^H\mathbf{H}_C)}{\log \rho}, \ j=1,\cdots,n_T
\label{eq:mu_def}
\end{align}
and $\mu_1 \ge \cdots \ge \mu_{n_T}$, we see that
$\sigma_j(\mathbf{H}_C)=\rho^{-\frac{1}{2}\mu_j}$, and from \eqref{eq:mr_appendix} that
\begin{align}
\sigma_i(\mathbf{M}) &\geq \rho^{\frac{1}{2}-\frac{rT}{\kappa}} \sigma_{\min}(\mathbf{G}) \sigma_{(i)}(\mathbf{I}_{T} \kron \mathbf{H}_R)) \nonumber\\ & \doteq \rho^{\frac{1}{2}-\frac{rT}{\kappa}} \sigma_{l_{2T}(i)}(\mathbf{H}_C) \nonumber\\ & = \rho^{\frac{-rT}{\kappa}+\frac{1}{2}(1-\mu_{l_{2T}(i)})},\label{eq:equ_sing_appendix1}\end{align}
where $l_T(i) \defeq \left\lceil {\frac{i}{T}}\right\rceil$, and where the asymptotic equality is due to the fact that $\sigma_{\min}(\mathbf{G}) \doteq \rho^0$.
Substituting from \eqref{eq:equ_sing_appendix1} in  \eqref{eq:equ_sing_appendix} we now have that
\begin{align}
\sigma_i(\mathbf{R}_{k})\dotgeq \rho^{\frac{-rT}{\kappa}+\frac{1}{2}(1-\mu_{l_{2T}(i)})^+}, \ \ i=1,\cdots,\kappa.
\label{eq:mu_heq}
\end{align}
Corresponding to \eqref{eq:nk11} we see that
\[\left[\sqrt{2k}+\frac{2\xi}{\sigma_i(\mathbf{R}_{k})}\right] \dotleq \rho^{\left(\frac{rT}{\kappa}-\frac{1}{2}(1-\mu_{l_{2T}(i)})^+\right)^+},\]
for any $i=1,\cdots,2n_TT$, and from \eqref{eq:nk11} we have that \begin{align}
N_k(\boldsymbol{\mu}) \dotleq \rho^{\sum_{i=1}^{k}{\left(\frac{rT}{\kappa}-\frac{1}{2}(1-\mu_{l_{2T}(i)})^+\right)^+}},
\label{eq:nk2}
\end{align}
where $\boldsymbol{\mu} = (\mu_1,\cdots,\mu_{n_T})$.
It follows that
\begin{align}
N_{SD}(\boldsymbol{\mu}) = \sum_{k=1}^{\kappa} N_k(\boldsymbol{\mu}) & \dotleq  \sum_{k=1}^{\kappa} \rho^{\sum_{i=1}^{k}{\left(\frac{rT}{\kappa}-\frac{1}{2}(1-\mu_{l_{2T}(i)})^+\right)^+}} \nonumber\\
& \doteq \rho^{\sum_{i=1}^{\kappa}{\left(\frac{rT}{\kappa}-\frac{1}{2}(1-\mu_{l_{2T}(i)})^+\right)^+}}\nonumber
\\
& \doteq \rho^{T \sum_{j=1}^{n_T} \left(\frac{r}{n_T}-(1-\mu_j)^+\right)^+},
\label{eq:n1}
\end{align}
where the last asymptotic equality is due to the multiplicity of the singular values.

Now consider the set \begin{align} \mathcal{T}(x)\defeq \left\{ \boldsymbol{\mu} \ | \ T \sum_{j=1}^{n_T} \left(\frac{r}{n_T}-(1-\mu_j)^+\right)^+ \geq x \right\}, \label{eq:mu_set_def} \end{align} and note that for any $y<x$, then \eqref{eq:n1} and $\boldsymbol{\mu} \notin  \mathcal{T}(y)$ jointly imply that $N_{SD} < \rho^x$, which in turn implies that $\prob{\boldsymbol{\mu} \notin  \mathcal{T}(y)} \leq \prob{N_{SD} < \rho^x}$ and consequently that %$\prob{\boldsymbol{\mu} \in  \mathcal{T}(x)} \geq \prob{N_{SD} \geq \rho^x}$. %For any $y < x$, we have that $\mathcal{T}(x) \subseteq \mathcal{T}(y)$ and it follows that
\begin{align}
\label{eq:set_1}
- \lim_{\rho \rightarrow \infty} \frac{\log \prob{N_{SD} \geq \rho^{x}}}{\log \rho} \geq -\lim_{\rho \rightarrow \infty} \frac{\log \prob{\boldsymbol{\mu} \in  \mathcal{T}(y)}}{\log \rho}.
\end{align}
In evaluating the right hand side of \eqref{eq:set_1} we note that $\mathcal{T}(y)$ is a closed set and thus, applying the large deviation principle~(cf. \cite{DZ:98}), we have that
\begin{align}
\label{eq:set_2}
-\lim_{\rho \rightarrow \infty} \frac{\log \prob{\boldsymbol{\mu} \in  \mathcal{T}(y)}}{\log \rho} \geq \inf_{\boldsymbol{\mu} \in  \mathcal{T}(y)}I(\boldsymbol{\mu})
\end{align}
for some rate function $I(\boldsymbol{\mu})$.  Consequently from \eqref{eq:set_1} and \eqref{eq:set_2}, it follows that
\begin{align}
\label{eq:set_3}
- \lim_{\rho \rightarrow \infty} \frac{\log \prob{N_{SD} \geq  \rho^{x}}}{\log \rho} \geq \inf_{\boldsymbol{\mu} \in  \mathcal{T}(y)}I(\boldsymbol{\mu}).
\end{align}
This lower bound specified in \eqref{eq:set_3} holds for any $y < x$.  Consequently to get the tightest possible bound, we need to find $\sup_{y<x}\inf_{\boldsymbol{\mu} \in  \mathcal{T}(y)}I(\boldsymbol{\mu})$. As $\inf_{\boldsymbol{\mu} \in  \mathcal{T}(y)}I(\boldsymbol{\mu})$ is non-decreasing and left-continuous in $y$, it follows that
\[\sup_{y<x}\inf_{\boldsymbol{\mu} \in  \mathcal{T}(y)}I(\boldsymbol{\mu})=\inf_{\boldsymbol{\mu} \in  \mathcal{T}(x)}I(\boldsymbol{\mu}).\] Consequently
\begin{align}
\label{eq:set_4}
- \lim_{\rho \rightarrow \infty} \frac{\log \prob{N_{SD} \geq  \rho^{x}}}{\log \rho} \geq \inf_{\boldsymbol{\mu} \in  \mathcal{T}(x)}I(\boldsymbol{\mu}),
\end{align}
which in conjunction with~\eqref{eq:complexity_exp2} gives that
\begin{align}
\label{eq:upper_bound_def}
c_{r-ld}(r) \leq \overline{c}_{r-ld}(r) &\defeq \inf \{x| \inf_{\boldsymbol{\mu} \in  \mathcal{T}(x)}I(\boldsymbol{\mu})>d_L(r)\} \nonumber \\
&{=}\sup \{x| \inf_{\boldsymbol{\mu} \in  \mathcal{T}(x)}I(\boldsymbol{\mu})\leq d_L(r)\} \nonumber \\
&{=}\max \{x| \inf_{\boldsymbol{\mu} \in  \mathcal{T}(x)}I(\boldsymbol{\mu})\leq d_L(r)\}
\end{align}
where the above follows from the aforementioned fact that $-\lim\limits_{\rho \rightarrow \infty} \frac{\log \prob{N_{SD} \geq  \rho^{x}}}{\log \rho}$ (and by extension also $\inf_{\boldsymbol{\mu} \in  \mathcal{T}(x)}I(\boldsymbol{\mu})$) is continuous and nondecreasing in $x$, and from the fact that $\mathcal{T}(x)$ is a closed set. Consequently $\overline{c}_{r-ld}(r)$ takes the form
\begin{subequations}  \label{eq:upper_bound_def1}
\begin{align}
\overline{c}_{r-ld}(r) \defeq \max_{\boldsymbol{\mu}} \; &  x \label{eq:obj_x}\\
\text{s.t.} \ \  & T{\sum_{j=1}^{n_T}{ \left(\frac{r}{n_T}-(1-\mu_j)^+\right)^+}} \geq x, \label{eq:const1}\\
& I(\boldsymbol{\mu}) \leq d_L(r),\label{eq:const2} \\
& \mu_1 \geq \cdots \geq \mu_{n_T} \geq 0. \label{eq:const3}
\end{align}
\end{subequations}
Furthermore since $\mathcal{T}(x)$ is a closed set, the maximum $x$ in \eqref{eq:upper_bound_def1} must be such that \eqref{eq:const1} is satisfied with equality, in which case $\overline{c}_{r-ld}(r)$ can be obtained as the solution to a constrained maximization problem according to
\begin{subequations}  \label{eq:cr_upper_bound_general}
\begin{align}
\overline{c}_{r-ld}(r) \defeq \max_{\boldsymbol{\mu}} \; & T{\sum_{j=1}^{n_T}{ \left(\frac{r}{n_T}-(1-\mu_j)^+\right)^+}} \label{eq:obj_cr_general}\\
\text{s.t.} \ \  & I(\boldsymbol{\mu}) \leq d_L(r),\label{eq:const1_cr_general} \\
& \mu_1 \geq \cdots \geq \mu_{n_T} \geq 0. \label{eq:const2_cr_general}
\end{align}
\end{subequations}

Equivalently for $\boldsymbol{\mu}^*=(\mu_1^*,\cdots,\mu_{n_T}^*)$ being one of the maximizing vectors\footnote{In general, \eqref{eq:cr_upper_bound_general} does not have a unique optimal point because $(a)^+$ is constant in $a$ for $a \leq 0$.}, i.e., such that $\boldsymbol{\mu}^* \in  \mathcal{T}(x)$ and $ I(\boldsymbol{\mu}^*) = d_L(r)$, then $\overline{c}_{r-ld}(r)$ takes the form
\begin{equation}
\overline{c}_{r-ld}(r)=T \sum_{j=1}^{n_T}{\left(\frac{r}{n_T}-(1-\mu^*_j)^+\right)^+}.
\label{eq:cr_upper_bound_point}
\end{equation}

As we will now show, the above bound is also shared by the ML-based sphere decoder, with or without MMSE preprocessing, irrespective of the full-rate code and the fading statistics.  Directly from \cite[Theorem 2]{JE:11}, and taking into consideration that MMSE-preprocessed lattice decoding is DMT optimal for any code~\cite{JE:10}, we recall that the equivalent upper bound for the ML-based sphere decoder, without MMSE preprocessing, takes the form
\begin{subequations}  \label{eq:cr_upper_bound_ml}
\begin{align}
\overline{c}_{ml}(r) \defeq \max_{\boldsymbol{\mu}} \; & T{\sum_{j=1}^{n_T}{ \min \left(\frac{r}{n_T}-1+\mu_j, \frac{r}{n_T}\right)^+}} \label{eq:obj_cr_ml}\\
\text{s.t.} \ \  & I(\boldsymbol{\mu}) \leq d_{L}(r),\label{eq:const1_cr_ml} \\
& \mu_1 \geq \cdots \geq \mu_{n_T} \geq 0\label{eq:const2_cr_ml}.
\end{align}
\end{subequations}
Comparing \eqref{eq:cr_upper_bound_general} and \eqref{eq:cr_upper_bound_ml} we are able to conclude that both the objective functions \eqref{eq:obj_cr_general} and \eqref{eq:obj_cr_ml} as well as both pairs of constraints are identical. To see this, we first note that for $0 \leq \mu_j \leq 1$, then
\begin{align*}
\min \left(\frac{r}{n_T}-1+\mu_j, \frac{r}{n_T}\right)^+ &=& \left(\frac{r}{n_T}-1+\mu_j\right)^+, \\ \left(\frac{r}{n_T}-(1-\mu_j)^+\right)^+ & = & \left(\frac{r}{n_T}-1+\mu_j\right)^+ ,\end{align*} and furthermore we note that for $\mu_j > 1$, then \[\min \left(\frac{r}{n_T}-1+\mu_j, \frac{r}{n_T}\right)^+ = \left(\frac{r}{n_T}-(1-\mu_j)^+\right)^+ = \frac{r}{n_T},\] which proves that $\overline{c}_{ml}(r)$ and $\overline{c}_{r-ld}(r)$ are identical.

In considering the case of MMSE-preprocessed ML SD, it is easy to see that the summands in the objective function in \eqref{eq:obj_cr_ml} will be modified to take the form $\min \left(\frac{r}{n_T}-(1-\mu_j)^+, \frac{r}{n_T}\right)^+ $ which can be seen to match $\eqref{eq:obj_cr_general}$ for all $\mu_j \geq 0$, which in turn concludes the proof that the upper bound $\overline{c}_{r-ld}(r)$ for MMSE-preprocessed lattice SD is also shared by the ML-based sphere decoder, with or without MMSE preprocessing, irrespective of the full-rate code, and for all fade statistics represented by monotonic rate functions.

\subsection{Lower bound on complexity of regularized lattice SD\label{app:lowerBoundCr}}
We will here, under the extra assumptions of regular i.i.d. fading statistics and of layered codes with natural decoding order, provide a lower bound that matches the upper bound in \eqref{eq:cr_upper_bound_point}.  The same bound and tightness will also apply to any full-rate code, under the assumption of a fixed, worst case decoding ordering.

The goal here is to show that at layer $k=2qT$, for some $q\in [1,\nt]$, the sphere decoder visits close to $\rho^{\overline{c}_{r-ld}(r)}$ nodes with a probability that is large compared to the probability of decoding error $\prob{\mathbf{s}_{L} \neq \mathbf{s}} \doteq \rho^{-d_L(r)}$, which from the expression of the complexity exponent~\eqref{eq:complexity_exp2}, will prove that $c_{r-ld}(r) = \overline{c}_{r-ld}(r)$.

Going back to \eqref{eq:cr_upper_bound_point}, we let $q$ be the largest integer for which
\begin{align}
\frac{r}{n_T}-(1-\mu^*_q)^+ >0,
\label{eq:q_def}
\end{align}
in which case \eqref{eq:cr_upper_bound_point} takes the form
\begin{align}
\overline{c}_{r-ld}(r)=T \sum_{j=1}^{q}{\frac{r}{n_T}-(1-\mu^*_j)^+}.
\label{eq:new_upperbound_def}
\end{align}
We recall from \eqref{eq:mu_def} that $\mu_j = -\frac{\log \sigma_j(\mathbf{H}_C^H\mathbf{H}_C)}{\log \rho}, \ j=1,\cdots,n_T$, and that $\boldsymbol{\mu}^*\in  \mathcal{T}(x)$ satisfies $ I(\boldsymbol{\mu}^*) = d_L(r)$ and maximizes \eqref{eq:obj_cr_general}.
We also note that without loss of generality we can assume that $q \geq 1$ as otherwise $\overline{c}_{r-ld}(r)=0$ (cf. \eqref{eq:cr_upper_bound_point}). Consequently it is the case that $\mu_j^* >0$ for $j=1,\cdots,q$.  Furthermore given the monotonicity of the rate function $I(\boldsymbol{\mu})$, and the fact that the objective function in \eqref{eq:cr_upper_bound_general} does not increase in $\mu_j$ beyond $\mu_j = 1$, we may also assume without loss of generality that $\mu_j^*  \leq 1$ for $j=1,\cdots,n_T$.

As in~\cite{JE:11} we proceed to define two events $\Omega_1$ and $\Omega_2$ which we will prove to be jointly sufficient so that, at layer $k=2qT$, the sphere decoder visits close to $\rho^{\overline{c}_{r-ld}(r)}$ nodes. These are given by
\begin{equation}
    \begin{split}
    \Omega_1 &\defeq  \{\mu_j^*-2\delta <\mu_j < \mu_j^*-\delta, j=1,\cdots,q \\
  & \quad \quad 0<\mu_j<\delta, j=q+1,\cdots,n_T\},
  \end{split}
\label{eq:omega_1}
\end{equation}
for a given small $\delta >0$, and
\begin{equation}
    \Omega_2 \defeq \{\sigma_1\left((\mathbf{I}_{T} \kron \mathbf{V}_p^H)\mathbf{G}_{|p}\right) \geq u\},
\label{eq:omega_2}
\end{equation}
for some given $u>0$, where for $p \defeq n_T-q$ then $\mathbf{G}_{|p}$ denotes the first $2pT$ columns of $\mathbf{G}$, and where $\mathbf{V}_p$ denotes the last $2p$ columns of $\mathbf{V}$ obtained by applying the singular value decomposition on $\mathbf{H}_R$, i.e., $\mathbf{H}_R=\mathbf{U}\mathbf{\Sigma}\mathbf{V}^H$, where \[\mathbf{\Sigma} \defeq \diag{\{\sigma_1(\mathbf{H}_R),\cdots, \sigma_{2n_T}(\mathbf{H}_R)\}}\] with $\sigma_1(\mathbf{H}_R)\leq \cdots \leq \sigma_{2n_T}(\mathbf{H}_R)$ and $\mathbf{V}\mathbf{V}^H=\mathbf{I}$. Hence, $\mathbf{V}_p^H$ corresponds to the $2p$ largest singular values of $\mathbf{H}_R$.

Note also that by choosing $\delta$ sufficiently small, and using the fact that $\mu_i^* >0$ for $i=1,\cdots,q$, we may without loss of generality assume that $\Omega_1$ implies that $\mu_j>0$ for all $j=1,\cdots,n_T$.

Modifying the approach in \cite[Theorem 1]{JE:11} to account for MMSE preprocessing and unconstrained decoding, the lower bound on the number of nodes visited at layer $k$ by the sphere decoder, is given by
\begin{align}
N_k \geq \prod_{i=1}^{k}\left[\frac{2\xi}{\sqrt{k}\sigma_i(\mathbf{R}_{k})}-\sqrt{k}\right]^+.
\label{eq:nk22}
\end{align}
In the following, and up until \eqref{eq:upperBound_sigmaRk}, we will work towards upper bounding $\sigma_i(\mathbf{R}_{k})$ so that we can then lower bound $N_k$.

Towards this let $$\mathbf{M}^{reg}_{|p} \defeq \left[\begin{array}{l}
\rho^{\frac{1}{2}-\frac{rT}{\kappa}}\mathbf{H}\mathbf{G}_{|p}\\ \alpha_r\mathbf{I}_{|p} \\ \end{array}\right] \in \mathbb{R}^{2(n_R+n_T)T \times 2pT}$$ contain the first $2pT$ columns of $\mathbf{M}^{reg}$ from~\eqref{eq:Mreg}, and note that
\[(\mathbf{M}^{reg}_{|p})^H\mathbf{M}^{reg}_{|p} = \rho^{1-\frac{2rT}{\kappa}}\mathbf{G}_{|p}^H\mathbf{H}^H\mathbf{H}\mathbf{G}_{|p} + \alpha^2_r\mathbf{I}\ ,\]
and that from \eqref{eq:h_appendix} we get
\begin{align*}
(\mathbf{M}^{reg}_{|p})^H\mathbf{M}^{reg}_{|p} = \rho^{1-\frac{2rT}{\kappa}}\mathbf{G}_{|p}^H(\mathbf{I}_{T} \kron \mathbf{H}_R^H\mathbf{H}_R)\mathbf{G}_{|p} + \alpha^2_r\mathbf{I} .
\end{align*}
Since
\begin{align*}
\mathbf{H}_{R}^H\mathbf{H}_{R} &= \mathbf{V}(\diag{\{\sigma_1(\mathbf{H}_{R}^H\mathbf{H}_{R}),\cdots, \sigma_{2n_T}(\mathbf{H}_{R}^H\mathbf{H}_{R})\}})\mathbf{V}^H \\
&= \mathbf{V}(\diag{\{\sigma_1(\mathbf{H}_{R}^H\mathbf{H}_{R}),\cdots, \sigma_{2n_T}(\mathbf{H}_{R}^H\mathbf{H}_{R})\}}\\
&\quad -\sigma_{(2q+1)}(\mathbf{H}_{R}^H\mathbf{H}_{R})\diag{\{\underbrace{0,\cdots,0}_{2q},\underbrace{1,\cdots,1}_{2p}\}})\mathbf{V}^H \\
&+\sigma_{(2q+1)}(\mathbf{H}_{R}^H\mathbf{H}_{R})\mathbf{V}(\diag{\{\underbrace{0,\cdots,0}_{2q},\underbrace{1,\cdots,1}_{2p}\}})\mathbf{V}^H,
\end{align*}
we have that
\begin{align*}
\mathbf{H}_{R}^H\mathbf{H}_{R} &\succeq \sigma_{(2q+1)}(\mathbf{H}_{R}^H\mathbf{H}_{R})\mathbf{V}(\diag{\{\underbrace{0,\cdots,0}_{2q},\underbrace{1,\cdots,1}_{2p}\}})\mathbf{V}^H \\
&=\sigma_{(2q+1)}(\mathbf{H}_{R}^H\mathbf{H}_{R})\mathbf{V}(\diag{\{\underbrace{0,\cdots,0}_{2q},\underbrace{1,\cdots,1}_{2p}\}})\\
& \quad \quad (\diag{\{\underbrace{0,\cdots,0}_{2q},\underbrace{1,\cdots,1}_{2p}\}})\mathbf{V}^H \\
&=\sigma_{(2q+1)}(\mathbf{H}_{R}^H\mathbf{H}_{R})\mathbf{V}_p\mathbf{V}_p^H
\end{align*}
where the last equality follows from the fact that $\mathbf{V}_p$ contains the last $2p$ columns of $\mathbf{V}$ and where $\mathbf{A} \succeq \mathbf{B}$ denotes that $\mathbf{A} - \mathbf{B}$ is positive-semidefinite. Since $\sigma_i(\mathbf{H}^H\mathbf{H}) \in \mathbb{R}$ and since the Kronecker product induces singular value multiplicity, it follows that
\begin{eqnarray*}
&&(\mathbf{M}^{reg}_{|p})^H\mathbf{M}^{reg}_{|p} \\
&\succeq&\rho^{1-\frac{2rT}{\kappa}}\sigma_{(2q+1)}(\mathbf{H}_{R}^H\mathbf{H}_{R})\mathbf{G}_{|p}^H(\mathbf{I}_{T} \kron \mathbf{V}_p\mathbf{V}_p^H)\mathbf{G}_{|p} + \alpha^2_r\mathbf{I}.
\end{eqnarray*}
With respect to the smallest singular value of $(\mathbf{M}^{reg}_{|p})^H\mathbf{M}^{reg}_{|p}$ we have
\begin{multline*}
\sigma_1((\mathbf{M}^{reg}_{|p})^H\mathbf{M}^{reg}_{|p})
\geq \rho^{1-\frac{2rT}{\kappa}}\sigma_{(2q+1)}(\mathbf{H}_{R}^H\mathbf{H}_{R}) \ \cdot \\
\sigma_1\left(\mathbf{G}_{|p}^H(\mathbf{I}_{T} \kron \mathbf{V}_p\mathbf{V}_p^H)\mathbf{G}_{|p}\right)+ \alpha^2_r
\end{multline*}
and consequently, given that $\mathbf{H}_{R} \in \Omega_2$, we have that
\begin{align}
\sigma_1(\mathbf{M}^{reg}_{|p}) &\geq \rho^{-\frac{rT}{\kappa}}\sqrt{u^2\rho\sigma_{l_2(2q+1)}(\mathbf{H}_{C}^H\mathbf{H}_{C})+ 1} \nonumber \\
&\doteq \rho^{-\frac{rT}{\kappa}}\rho^{\frac{1}{2}(1-\mu_{q+1})^+} \quad \nonumber \\
&\geq \rho^{-\frac{rT}{\kappa}+\frac{1}{2}(1-\delta)^+},
\label{eq:sig_21}
\end{align}
where the first inequality follows from \eqref{eq:omega_2}, the exponential equality follows from \eqref{eq:mu_def} and from the fact that $u>0$ is fixed and independent of $\rho$, and the last inequality follows from \eqref{eq:omega_1}.

From \eqref{eq:mr_appendix} we have that
\begin{eqnarray}
\sigma_i(\mathbf{M}^{reg})& {\leq}& \rho^{\frac{-rT}{\kappa}}  \sqrt{(1+\rho (\sigma_{\kappa}(\mathbf{G})\sigma_{l_{2T}(i)}(\mathbf{H}_{C}))^2)}\nonumber\\
&{\doteq}& \rho^{\frac{-rT}{\kappa}+\frac{1}{2}(1-\mu_{l_{2T}(i)})^+}, \ \ i=1,\cdots,2n_TT, \ \ \ \ \
\label{eq:sig_20_1}
\end{eqnarray}
where the asymptotic equality follows from the fact that $\sigma_{\kappa}(\mathbf{G})$ is fixed and independent of $\rho$. Furthermore \eqref{eq:omega_1} gives that for $i=1,\cdots,2qT$ then
\begin{align}
\sigma_i(\mathbf{M}^{reg}) \dotleq \rho^{-\frac{rT}{\kappa}+\delta+\frac{1}{2}(1-\mu^*_{l_{2T}(i)})^+},
\label{eq:sig_20}
\end{align}
where we have made use of the fact that $\mu^*_j \leq 1$ for $j=1,\cdots,n_T$.

Given that $\mu^*_j >0$ for $j=1,\cdots,q$, then for sufficiently small $\delta$ and for $i=1,\cdots,2qT$, we have that \[-\frac{rT}{\kappa}+\frac{1}{2}(1-\delta)^+ \geq -\frac{rT}{\kappa}+\delta+\frac{1}{2}(1-\mu^*_{l_{2T}(i)})^+ ,\] which means that for sufficiently small $\delta$, a comparison of \eqref{eq:sig_21} and \eqref{eq:sig_20} yields\[\sigma_i(\mathbf{M}^{reg}) < \sigma_1(\mathbf{M}^{reg}_{|p}),\]
for $i=1,\cdots,2qT$.  The above inequality allows us to apply Lemma~3 in \cite{JE:11}, which in turn gives that
\begin{align}
\sigma_i(\mathbf{R}_{k}) \leq \left[\frac{\sigma_\kappa(\mathbf{M}^{reg})}{\sigma_1(\mathbf{M}^{reg}_{|p})}+1\right]\sigma_i(\mathbf{M}^{reg}),
\label{eq:lower_sig_1}
\end{align}
for $i=1,\cdots,2qT$.

Setting $i=\kappa$ in~\eqref{eq:sig_20_1} upper bounds the maximum singular value of $\mathbf{M}^{reg}$ as
\begin{align}
\sigma_\kappa(\mathbf{M}^{reg}) \dotleq \rho^{-\frac{rT}{\kappa}+\frac{1}{2}(1-\mu_{n_T})^+} \leq \rho^{\frac{1}{2}-\frac{rT}{\kappa}},
\label{eq:sigma_m_max}
\end{align}
where the last inequality is due to the fact that $\mu_j \geq 0$. Consequently combining \eqref{eq:sigma_m_max} and \eqref{eq:sig_21} gives that \[\left[\frac{\sigma_\kappa(\mathbf{M}^{reg})}{\sigma_1(\mathbf{M}^{reg}_{|p})}+1\right] \dotleq \rho^{\frac{1}{2}\delta},\]
which together with \eqref{eq:sig_20} and \eqref{eq:lower_sig_1} gives that
\begin{align} \label{eq:upperBound_sigmaRk}
\sigma_i(\mathbf{R}_{k}) \dotleq \rho^{-\frac{rT}{\kappa}+\frac{3}{2}\delta+\frac{1}{2}(1-\mu^*_{l_{2T}(i)})^+}, \ \ i=1,\cdots,2qT.\end{align}  Consequently, going back to \eqref{eq:nk22}, we have that
\begin{align}
\left[\frac{2\xi}{\sqrt{k}\sigma_i(\mathbf{R}_{k})}-\sqrt{k}\right]^+ \dotgeq \rho^{\left(\frac{rT}{\kappa}-\frac{3}{2}\delta-\frac{1}{2}(1-\mu^*_{l_{2T}(i)})^+\right)} >0
\label{eq:nk_19}
\end{align}
and furthermore for $i=1,\cdots,2qT$, we have that $\frac{rT}{\kappa}-\frac{3}{2}\delta-\frac{1}{2}(1-\mu^*_{l_{2T}(i)})^+ > 0$ directly from definition of $q$ and for sufficiently small $\delta$. As a result, for $k \leq 2qT$ we have that
\begin{align}
N_k &\dotgeq \prod_{i=1}^{k}\rho^{\left(\frac{rT}{\kappa}-\frac{3}{2}\delta-\frac{1}{2}(1-\mu^*_{l_{2T}(i)})\right)} \\ &=\rho^{\sum_{i=1}^{k}{\left(\frac{rT}{\kappa}-\frac{1}{2}(1-\mu^*_{l_{2T}(i)})^+\right)}-\frac{3}{2}k\delta},
\label{eq:nk_18_1}
\end{align}
and setting $k=2qT$ we have that
\begin{align}
N_{2qT}&\dotgeq \rho^{\left(\sum_{i=1}^{2qT}{\left(\frac{rT}{\kappa}-\frac{1}{2}(1-\mu^*_{l_{2T}(i)})^+\right)}-3qT\delta \right)} \\
&=\rho^{\left(T \sum_{j=1}^{q}{\left(\frac{rT}{\kappa}-(1-\mu^*_j)^+\right)}-3qT\delta \right)}  \\
&=\rho^{(\overline{c}_{r-ld}(r) - 3qT\delta)},
\label{eq:nk_18}
\end{align}
where the last equality follows from \eqref{eq:new_upperbound_def}. Consequently
\[N_{SD} \geq N_{2qT} \dotgeq \rho^{\overline{c}_{r-ld}(r) - 3qT\delta},\] for small $\delta>0$. Given that $\delta$ can be chosen arbitrarily small, and given that events $\Omega_1$ and $\Omega_2$ occur, then the number of nodes visited by the SD at layer $2qT$ is arbitrarily close to the upper bound of $\rho^{\overline{c}_{r-ld}(r)}$.

Now to show that $c_{r-ld}(r)\geq \overline{c}_{r-ld}(r) - 3qT\delta$, we just have to prove that $-\displaystyle\lim_{\rho\to\infty}\frac{\prob{N_{SD} \dotgeq \rho^{\overline{c}_{r-ld}(r) - 3qT\delta}}}{\log \rho} < d_L(r)$. Toward this we note that as \eqref{eq:omega_1} and \eqref{eq:omega_2} imply that $N_{SD} \dotgeq \rho^{\overline{c}_{r-ld}(r) - 3qT\delta}$, it follows that
\[\prob{N_{SD} \dotgeq \rho^{\overline{c}_{r-ld}(r) - 3qT\delta}} \geq \prob{\Omega_1 \cap \Omega_2} =\prob{\Omega_1}\prob{\Omega_2}\]
where the equality follows from the i.i.d. assumption on the entries in $\mathbf{H}_C$, which makes the singular values of $\mathbf{H}_C^H\mathbf{H}_C$ independent of the singular vectors of $\mathbf{H}_C^H\mathbf{H}_C$ \cite{RSB:05},\cite{TV:04}, and which in turn also implies independence of the singular values of $\mathbf{H}_C^H\mathbf{H}_C$ (event $\Omega_1$) from the singular vectors of $\mathbf{H}_R^H\mathbf{H}_R$ (event $\Omega_2$).

We now turn to \cite[Lemma 2]{JE:11} and recall that for the layered codes assumed here, as well as for any full-rate design and some non-random fixed decoding ordering (corresponding to a permutation of the columns of $\bfG$), there exists a unitary matrix $\mathbf{V}^{'}_{\!\!p}$ such that $\rank{\left((\mathbf{I}_{T} \kron (\mathbf{V}^{'}_{\!\!p})^H)\mathbf{G}_{|p}\right)} =2pT$ i.e., that
\[\sigma_1\left((\mathbf{I}_{T} \kron (\mathbf{V}^{'}_{\!\!p})^H)\mathbf{G}_{|p}\right) > 0.\]
However, by continuity of singular values \cite{HJ:85} it follows for sufficiently small $u>0$ (cf.\eqref{eq:omega_2}) that $\prob{\Omega_2}>0$, which implies\footnote{In light of the fact that event $\mathbf{V}^{'}_{\!\!p}$ has zero measure, what the continuity of eigenvalues guarantees is that we can construct a neighborhood of matrices around $\mathbf{V}^{'}_{\!\!p}$ which are full rank, and which have a non zero measure.  We also note that the matrices $\mathbf{V}^{'}_{\!\!p}$ can be created recursively, starting from a single matrix $\mathbf{V}^{'}_{\!\!\nt}$.} that $\prob{\Omega_2} \doteq \rho^0$ as $\Omega_2$ is independent of $\rho$.  This in turn implies that
\begin{align} \label{eq:Pw1}
\prob{N_{SD} \dotgeq \rho^{\overline{c}_{r-ld}(r) - 3qT\delta}} \dotgeq \prob{\Omega_1}.
\end{align}

With $\Omega_1$ being an open set, we have that
\begin{eqnarray}
-\displaystyle\lim_{\rho\to\infty}\frac{\prob{\Omega_1}}{\log \rho} &\leq& \inf_{\boldsymbol{\mu} \in \Omega_1} I(\boldsymbol{\mu}), \nonumber \\
&{=} & \sum_{ j=1}^{q} (|n_T-n_R|+2 j-1)(\mu_{ j}^*-2\delta), \nonumber \\
&=&d_L(r)-2(|n_T-n_R|+q)q\delta, \nonumber \\
&< &d_L(r),
\label{eq:prob_lower1}
\end{eqnarray}
where the above follows from the monotonicity of the rate function
\[I(\boldsymbol{\mu}) = \sum_{j=1}^{n_T} (|n_T-n_R|+2j-1)\mu_i + \frac{{n_R}{n_T}t}{2}\mu_{n_T},\] evaluated at \[ \{\mu_1^*-2\delta\, \cdots ,\mu_q^*-2\delta, 0, \cdots ,0\}  = \arg\inf_{\boldsymbol{\mu} \in \Omega_1} I(\boldsymbol{\mu})  ,\]and\footnote{Recall that parameter $t$ was previously introduced as a parameter that regulates the near zero behavior of the random variable.} also follows from the fact that, by definition, $I(\boldsymbol{\mu}^*)=d_L(r)$.

Consequently from \eqref{eq:Pw1} we have that
\begin{eqnarray}
-\displaystyle\lim_{\rho\to\infty}\frac{\prob{N_{SD} \dotgeq \rho^{\overline{c}_{r-ld}(r) - 3qT\delta}}}{\log \rho} &<& d_L(r),
\label{eq:prob_lower}
\end{eqnarray}
and directly from the definition of the complexity exponent, we have that $c_{r-ld}(r) \geq \overline{c}_{r-ld}(r)-3qT\delta$. As the bound holds for arbitrarily small $\delta>0$, it follows that $c_{r-ld}(r)=\overline{c}_{r-ld}(r)$. Directly from \cite[Theorem 4]{JE:11} which analyzes the ML-based complexity exponent $c_{ml}(r)$, together with the fact that the ML-based sphere decoder, with or without MMSE preprocessing, shares the same upper bound $\overline{c}_{r-ld}(r)$ as the MMSE-preprocessed lattice decoder, gives that $c_{ml}(r)=\overline{c}_{r-ld}(r)$, which in turns implies that \[c_{r-ld}(r)=c_{ml}(r).\]
This establishes Theorem~\ref{thm:cr_mmse} and Corollary~\ref{thm:cr_mmse_ordering}.  $\square$

%%%%%%%%%%%%%%%%%%%%%%%%%%%%%%%%%%%%%%%%%%%%%%%%%%%%%%%%%%%%%%%%%%%%%%%%%%%%%%%%%%%%%%%%%%%%%%%%%%%
%%%%%%%%%%%%%%%%%%%%%%%%%%%%%%%%%%%%%%%%%%%%%%%%%%%%%%%%%%%%%%%%%%%%%%%%%%%%%%%%%%%%%%%%%%%%%%%%%%%

\section{Proof for Corollaries~\ref{thm:cr_mmse1}~and~\ref{thm:cr_mmse_bound} \label{app:corollaries}}
Section~\ref{app:Upper} shows that $\overline{c}_{r-ld}(r)$ can be obtained as the solution to the constrained maximization problem
\begin{subequations}  \label{eq:cr_upper_bound_general2}
\begin{align}
\overline{c}_{r-ld}(r) \defeq \max_{\boldsymbol{\mu}} \; & T{\sum_{j=1}^{n_T}{ \left(\frac{r}{n_T}-(1-\mu_j)^+\right)^+}} \nonumber\\
\text{s.t.} \ \  & I(\boldsymbol{\mu}) \leq d_L(r),\label{eq:const1_cr_general2} \\
& \mu_1 \geq \cdots \geq \mu_{n_T} \geq 0. \label{eq:const2_cr_general2}
\end{align}
\end{subequations}
In some cases though, further knowledge of the error performance of the encoder and decoder, can result in an explicit characterization of the complexity exponent.  Take for instance the case of DMT optimal encoding \cite{EKP+:06,ORB:06} and DMT optimal MMSE-preprocessed lattice decoding \cite{GCD:04,JE:10}, where the constraint $I(\boldsymbol{\mu}) \leq d_L(r)$ in~\eqref{eq:const1_cr_general2} reverts to the constraint $\sum_{j=1}^{n_T}(1-\mu_{j})^+ \geq r$ (cf.~\cite{JE:10}), which may be recognized to correspond to the no-outage region (cf. \cite{ZT:03}).  In this case $\overline{c}_{r-ld}(r)$ can then be explicitly obtained from the optimization problem
\begin{subequations}  \label{eq:cr_upper_bound_DMT_opt}
\begin{align}
\overline{c}_{r-ld}(r) = \max_{\boldsymbol{\mu}} \; & T {\sum_{j=1}^{n_T}{\left(\frac{r}{n_T}-(1-\mu_j)^+\right)^+}} \label{eq:obj_cr_DMT_opt} \\
\text{s.t.} \; & \sum_{j=1}^{n_T}(1-\mu_{j})^+ \geq r \label{eq:const1_cr_DMT_opt} \\
& \mu_1 \geq ....\geq \mu_{n_T} \geq 0, \label{eq:const2_cr_DMT_opt}
\end{align}
\end{subequations}
which can be solved in a straightforward manner to give that
\begin{align*}
\overline{c}_{r-ld}(r) = \frac{T}{n_T}\left(r(n_T-\left\lfloor{r}\right\rfloor-1) +(n_T\left\lfloor{r}\right\rfloor -r(n_T-1))^+\right),
\end{align*}
describing the upper bound on the complexity exponent for MMSE-preprocessed lattice sphere decoding of DMT optimal full-rate codes, which for minimum delay ($n_T=T$) DMT optimal full-rate codes takes the form
\begin{align}
\overline{c}_{r-ld}(r)=r(n_T-\left\lfloor{r}\right\rfloor-1) +(n_T\left\lfloor{r}\right\rfloor -r(n_T-1))^+,
\label{eq:upperbound_mmse_threaded}
\end{align}
and which further simplifies to
\[\overline{c}_{r-ld}(r) = r(n_T-r),\]
for integer multiplexing gains $r=0,1,\cdots,n_T$.  In conjunction with the lower bound in Section~\ref{app:lowerBoundCr}, under the conditions layered codes in Corollary~\ref{thm:cr_mmse1}, we have that $c_{r-ld}(r) = \overline{c}_{r-ld}(r)$, which proves Corollary~\ref{thm:cr_mmse1}. $\square$

Moving on to the universal upper bound, we can see from \eqref{eq:cr_upper_bound_general} that, regardless of the fading statistics and the corresponding $I(\boldsymbol{\mu})$, the exponent $\overline{c}_{r-ld}(r)$ is non-decreasing in $d_L(r)$ and is hence maximized when $d_L(r)$ is itself maximized, i.e., it is maximized in the presence of DMT optimal encoding and decoding.   Combined with the fact that the corresponding maximization problem in \eqref{eq:cr_upper_bound_DMT_opt} does not depend on the fading distribution, other than the natural fact that its tail must vanish exponentially fast, results in the fact that, for any full-rate code and statistical characterization of the channel, the complexity of MMSE-preprocessed lattice SD is universally upper bounded as (cf.\cite{JE:11})
\begin{align}
\frac{T}{n_T}\left(r(n_T-\left\lfloor{r}\right\rfloor-1) +(n_T\left\lfloor{r}\right\rfloor -r(n_T-1))^+\right).
\label{eq:upperbound_universal}
\end{align}
This proves Corollary \ref{thm:cr_mmse_bound}. $\square$

%%%%%%%%%%%%%%%%%%%%%%%%%%%%%%%%%%%%%%%%%%%%%%%%%%%%%%%%%%%%%%%%%%%%%%%%%%%%%%%%%%%%%%%%%%%%%%%%%%%%%%%%%%%%%%%%%%%%%%%%
%%%%%%%%%%%%%%%%%%%%%%%%%%%%%%%%%%%%%%%%%%%%%%%%%%%%%%%%%%%%%%%%%%%%%%%%%%%%%%%%%%%%%%%%%%%%%%%%%%%%%%%%%%%%%%%%%%%%%%%%

\section{Proof for Lemma~\ref{lem:lambdamin}}
\label{app:lambdamin}
For $\mathbf{R}_r^H\mathbf{R}_r=\mathbf{M}_r^H\mathbf{M}_r+\alpha_r^2\mathbf{I}$ (cf.~\eqref{eq:feedback})\footnote{Note the transition to the notation reflecting the dependence of $\mathbf{R}$ on $r$.}, it follows by the bounded orthogonality defect of LLL reduced bases that there is a constant $K_{\kappa} >0$ independent of $\mathbf{R}_r$ and $\rho$, for which~(cf.~\cite{LLL:82} and the proof in \cite{TMK:07})
\begin{align}
\sigma_{max}({\mathbf{\tilde{R}}_r}^{-1}) \leq \frac{K_{\kappa}}{\lambda(\mathbf{R}_r)}
\label{eq:singularitymax}
\end{align}
where
\begin{eqnarray}
\lambda(\mathbf{R}_r) \defeq \min_{\mathbf{c}\in \mathbb{Z}^{\kappa} \backslash \mathbf{0}} \|\mathbf{R}_r\mathbf{c}\|
\label{eq:lambdar}
\end{eqnarray}
denotes the shortest vector in the lattice generated by $\mathbf{R}_r$.
As a result we have that
\begin{align}
\sigma_{min}(\mathbf{\tilde{R}}_r) \geq \frac{\lambda(\mathbf{R}_r)}{K_{\kappa}}.
\label{eq:singularitymin}
\end{align}
Looking to lower bound $\sigma_{min}(\mathbf{\tilde{R}}_{r})$, we seek a bound on $\lambda(\mathbf{R}_{r})$.  Towards this let $r'=r-\gamma$ for some $r \geq \gamma>0$, in which case for $\mathbf{s}$ being the transmitted symbol vector, and for any $\mathbf{\hat{s}} \in \mathbb{Z}^{\kappa}$ such that $\mathbf{\hat{s}} \neq \mathbf{s}$, it follows that
\begin{eqnarray}
\|\mathbf{r}-\mathbf{R}_{r'}\mathbf{\hat{s}}\| &=& \|(\mathbf{r}-\mathbf{R}_{r'}\mathbf{s})+\mathbf{R}_{r'}(\mathbf{s}-\mathbf{\hat{s}})\|  \nonumber \\
&\leq&\|(\mathbf{r}-\mathbf{R}_{r'}\mathbf{s})\|+\|\mathbf{R}_{r'}(\mathbf{s}-\mathbf{\hat{s}})\|
\label{eq:rc1}
\end{eqnarray}
and
\begin{eqnarray}
\|\mathbf{R}_{r'}(\mathbf{s}-\mathbf{\hat{s}})\| &\geq& \|\mathbf{r}-\mathbf{R}_{r'}\mathbf{\hat{s}}\| - \|(\mathbf{r}-\mathbf{R}_{r'}\mathbf{s})\| \nonumber \\
&=& \|\mathbf{r}-\mathbf{R}_{r'}\mathbf{\hat{s}}\| - \|\mathbf{w}\|.
\label{eq:rc2}
\end{eqnarray}
From \eqref{eq:rc2} it is clear that to find a lower bound on $\lambda(\mathbf{R}_{r'})$, we need to lower bound $\|\mathbf{r}-\mathbf{R}_{r'}\mathbf{\hat{s}}\|$ for all $\mathbf{\hat{s}} \in \mathbb{Z}^{\kappa}$ and upper bound $\|\mathbf{w}\|$.  Let us, for now, assume that $\left\|\mathbf{w}\right\|^2 \leq \rho^{b}$.
To lower bound $\|\mathbf{r}-\mathbf{R}_{r'}\mathbf{\hat{s}}\|$, we draw from the equivalence of MMSE preprocessing and the regularized metric (cf. equation (45) in \cite{JE:10}), and rewrite
\begin{align}
\left\|\mathbf{r}-\mathbf{R}_{r'}\mathbf{\hat{s}}\right\|^2 = \left\|\mathbf{y}-\mathbf{M}_{r'}\mathbf{\hat{s}}\right\|^2 + \alpha_{r'}^2\left\|\mathbf{\hat{s}}\right\|^2 - c,
\label{eq:mmse_rld_eq1}
\end{align}
where $c \defeq \mathbf{y}^H[\mathbf{I}-\mathbf{M}_{r'}^H(\mathbf{M}_{r'}^H\mathbf{M}_{r'}+\alpha_{r'}^2\mathbf{I})^{-1}\mathbf{M}_{r'}]\mathbf{y} \geq 0$.
We now note that for $\mathbf{\hat{s}}=\mathbf{s}$ then $\left\|\mathbf{y}-\mathbf{M}_{r'}\mathbf{s}\right\|^2 +\alpha_{r'}^2\left\|\mathbf{s}\right\|^2 \dotleq \rho^{b}$, and since the left hand side of \eqref{eq:mmse_rld_eq1} cannot be negative, and furthermore given that $c$ is independent of $\mathbf{\hat{s}}$, we conclude that $c \dotleq \rho^{b}$.

We will now proceed to lower bound $\left\|\mathbf{y}-\mathbf{M}_{r'}\mathbf{\hat{s}}\right\|^2 + \alpha_{r'}^2 \left\|\mathbf{\hat{s}}\right\|^2$ and then use \eqref{eq:mmse_rld_eq1} to lower bound $\|\mathbf{r}-\mathbf{R}_{r'}\mathbf{\hat{s}}\|$.
Towards lower bounding $\left\|\mathbf{y}-\mathbf{M}_{r'}\mathbf{\hat{s}}\right\|^2 + \alpha_{r'}^2 \left\|\mathbf{\hat{s}}\right\|^2$ we draw from Theorem 1 in \cite{JE:10} and we let $\mathcal{B}$ be the spherical region given by \[\mathcal{B} \defeq \{d \in \mathbb{R}^{\kappa}|\left\|\mathbf{d}\right\|^2 \leq \Gamma^2 \}\]
where the radius $\Gamma > 0$ is independent of $\rho$ and is chosen so that $\mathbf{d}_1+\mathbf{d}_2 \in \mathcal{R}$ for any $\mathbf{d}_1, \mathbf{d}_2 \in \mathcal{B}$. The existence of the set $\mathcal{B}$ follows by the assumption that $\mathbf{0}$ is contained in the interior of $\mathcal{R}$. Now let \[\nu_{r'} \defeq \min_{\mathbf{d} \in \rho^{\frac{{r'}T}{\kappa}}\mathcal{B} \cap \mathbb{Z}^{\kappa}:\mathbf{d}\neq \mathbf{0}} \frac{1}{4}\left\|\mathbf{M}_{r'}\mathbf{d}\right\|^2,\]
and for given $\gamma > \zeta > 0$ choose $b >0$ such that
\[\frac{2\zeta T}{\kappa} >b>0.\]
This may clearly be done for arbitrary $\zeta >0$. We will in the following temporarily assume that $\nu_{r'+\zeta} \geq 1$ and prove that, together with $\left\|\mathbf{w}\right\|^2 \leq \rho^{b}$, the two conditions are sufficient for $\lambda(\mathbf{\tilde{R}}_{r'}) \dotgeq \rho^{\frac{\zeta T}{\kappa}}$ to hold.

In order to bound the metric for $\mathbf{\hat{s}} \in \mathbb{Z}^{\kappa}$ where $\mathbf{\hat{s}} \neq \mathbf{s}$, we note that $\nu_{r'+\zeta} \geq 1 $ implies that $\forall \mathbf{d} \in \rho^{\frac{(r'+\zeta) T}{\kappa}}\mathcal{B} \cap \mathbb{Z}^{\kappa}, \mathbf{d}\neq \mathbf{0}$ it is the case that
\begin{eqnarray}
\frac{1}{4}\left\|\mathbf{M}_{r'+\zeta}\mathbf{d}\right\|^2&\geq& 1  \nonumber \\
\frac{1}{4}\left\|\rho^{\frac{1}{2}-\frac{(r'+\zeta) T}{\kappa}}\mathbf{HG}\mathbf{d}\right\|^2 &\stackrel{(a)}{\geq}& 1  \nonumber \\
\frac{1}{4}\left\|\rho^{\frac{1}{2}-\frac{r'T}{\kappa}}\mathbf{HG}\mathbf{d}\right\|^2 &\geq& \rho^{\frac{2\zeta T}{\kappa}}  \nonumber
\label{eq:nvr1}
\end{eqnarray}
where $(a)$ follows from the fact that $\mathbf{M}_r= \rho^{\frac{1}{2}-\frac{rT}{\kappa}}\mathbf{H}\mathbf{G}$.  Consequently
\begin{align}
\frac{1}{4}\left\|\mathbf{M}_{r'}\mathbf{d}\right\|^2 \geq \rho^{\frac{2\zeta T}{\kappa}}, \ \forall  \mathbf{d} \in \rho^{\frac{(r'+\zeta) T}{\kappa}}\mathcal{B} \cap \mathbb{Z}^{\kappa}, \mathbf{d}\neq \mathbf{0}.
\label{eq:nvr2}
\end{align}
As $\mathcal{R}$ is bounded, and as $\zeta >0$, it holds that $\mathcal{R} \subset \frac{1}{2}\rho^{\frac{\zeta T}{\kappa}}\mathcal{B}$ for all $\rho \geq \rho_1$, for a sufficiently large $\rho_1$. This implies that $\mathbf{s} \in \frac{1}{2}\rho^{\frac{(r'+\zeta) T}{\kappa}}\mathcal{B}$ for $\rho \geq \rho_1$ since $\mathbf{s} \in \rho^{\frac{r'T}{\kappa}}\mathcal{R}$.

For $\mathbf{s}, \mathbf{d} \in \frac{1}{2}\rho^{\frac{(r'+\zeta) T}{\kappa}}\mathcal{B} \cap \mathbb{Z}^{\kappa}$, there exists an $\mathbf{\hat{s}} \in \rho^{\frac{(r'+\zeta) T}{\kappa}}\mathcal{B} \cap \mathbb{Z}^{\kappa}$, $\mathbf{\hat{s}} \neq \mathbf{s}$, such that $ \mathbf{\hat{s}}= \mathbf{d} + \mathbf{s}$. Hence for any $\mathbf{\hat{s}} \in \rho^{\frac{(r'+\zeta) T}{\kappa}}\mathcal{B} \cap \mathbb{Z}^{\kappa}$, we have from~\eqref{eq:nvr2} that
\begin{align}
\frac{1}{4}\left\|\mathbf{M}_{r'}(\mathbf{\hat{s}} - \mathbf{s})\right\|^2 = \frac{1}{4}\left\|\mathbf{M}_{r'}\mathbf{d}\right\|^2 \geq \rho^{\frac{2\zeta T}{\kappa}}.
\label{eq:nvr3}
\end{align}
As $\left\|\mathbf{w}\right\|^2 \leq \rho^{b}$, it follows that $\frac{1}{4}\left\|\mathbf{M}_{r'}\mathbf{d}\right\|^2 \geq \left\|\mathbf{w}\right\|^2 $ for large $\rho$, and that
\begin{align}
\left\|\mathbf{y}-\mathbf{M}_{r'}\mathbf{\hat{s}}\right\|^2 =\left\|\mathbf{M}_{r'}(\mathbf{s} - \mathbf{\hat{s}})+\mathbf{w}\right\|^2 \dotgeq \rho^{\frac{2\zeta T}{\kappa}}.
\label{eq:nvr4}
\end{align}
Consequently
\begin{align}
\left\|\mathbf{y}-\mathbf{M}_{r'}\mathbf{\hat{s}}\right\|^2 +\alpha_{r'}^2\left\|\mathbf{\hat{s}}\right\|^2 \dotgeq \rho^{\frac{2\zeta T}{\kappa}}.
\label{eq:nvr5}
\end{align}

On the other hand if $\mathbf{\hat{s}} \notin \rho^{\frac{(r'+\zeta) T}{\kappa}}\mathcal{B}$, then by definition of $\mathcal{B}$ we have that $\alpha_{r'}^2\left\|\mathbf{\hat{s}}\right\|^2 \geq \frac{1}{4}\Gamma^2 \rho^{\frac{2\zeta T}{\kappa}} $, and consequently that
\begin{align}
\left\|\mathbf{y}-\mathbf{M}_{r'}\mathbf{\hat{s}}\right\|^2 +\alpha_{r'}^2\left\|\mathbf{\hat{s}}\right\|^2 \geq \frac{1}{4}\Gamma^2 \rho^{\frac{2\zeta T}{\kappa}}.
\label{eq:nvr6}
\end{align}
From \eqref{eq:nvr5} and \eqref{eq:nvr6} we then conclude that
\begin{align}
\left\|\mathbf{y}-\mathbf{M}_{r'}\mathbf{\hat{s}}\right\|^2 +\alpha_{r'}^2\left\|\mathbf{\hat{s}}\right\|^2 \dotgeq \rho^{\frac{2\zeta T}{\kappa}}.
\label{eq:nvr7}
\end{align}
Given \eqref{eq:nvr5} and \eqref{eq:nvr7}, for any $\mathbf{\hat{s}} \in \mathbb{Z}^{\kappa}$ such that $\mathbf{\hat{s}} \neq \mathbf{s}$, it is the case that $\left\|\mathbf{y}-\mathbf{M}_{r'}\mathbf{\hat{s}}\right\|^2 +\alpha_{r'}^2\left\|\mathbf{\hat{s}}\right\|^2 \dotgeq \rho^{\frac{2\zeta T}{\kappa}}$, which combined with $c \dotleq \rho^{b}$ allows for \eqref{eq:mmse_rld_eq1} to give that
\begin{align}
\left\|\mathbf{r}-\mathbf{R}_{r'}\mathbf{\hat{s}}\right\|^2 \dotgeq \rho^{\frac{2\zeta T}{\kappa}}.
\label{eq:metric_mmse}
\end{align}
Applying \eqref{eq:lambdar} and \eqref{eq:rc2}, we have
\begin{eqnarray}
\lambda(\mathbf{R}_{r'}) &\geq& \|\mathbf{r}-\mathbf{R}_{r'}\mathbf{\hat{s}}\| - \|\mathbf{w}\|\nonumber \\
&\dotgeq & \rho^{\frac{\zeta T}{\kappa}} - \rho^{\frac{b}{2}} \nonumber \\
& \doteq & \rho^{\frac{\zeta T}{\kappa}}
\label{eq:rc3}
\end{eqnarray}
where the exponential inequality follows from \eqref{eq:metric_mmse}. Furthermore we know that
\begin{align}
\lambda(\mathbf{R}_{r})=\rho^{\frac{-\gamma T}{\kappa}}\lambda(\mathbf{R}_{r'}) \dotgeq \rho^{\frac{-\epsilon T}{\kappa}}
\label{eq:lam_rf}
\end{align}
where $\epsilon=\gamma-\zeta$, $r \geq \epsilon >0$, and from \eqref{eq:singularitymin} and \eqref{eq:lam_rf} it follows that $\sigma_{min}(\mathbf{\tilde{R}}_{r}) \dotgeq \rho^{\frac{-\epsilon T}{\kappa}}$.

We now note that the above implies that for $\nu_{r'+\zeta} \geq 1$ and $\left\|\mathbf{w}\right\|^2 \leq \rho^{b}$ then $\sigma_{min}(\mathbf{\tilde{R}}_{r}) \dotgeq \rho^{\frac{-\epsilon T}{\kappa}}$, and thus applying the union bound yields
\begin{eqnarray*}
\prob{\sigma_{min}(\mathbf{\tilde{R}}_{r}) \stackrel{.}{<} \rho^{\frac{-\epsilon T}{\kappa}}} &= &\prob{(\nu_{r'+\zeta} < 1) \cup (\left\|\mathbf{w}\right\|^2 > \rho^{b})} \\
& \leq & \prob{\nu_{r'+\zeta} < 1} + \prob{\left\|\mathbf{w}\right\|^2 > \rho^{b}}.
\label{eq:prob_e1}
\end{eqnarray*}

We know from the exponential tail of the Gaussian distribution that $\prob{\left\|\mathbf{w}\right\|^2 > \rho^{b}} \doteq \rho^{-\infty}$ and from Lemma 1 in \cite{JE:10} that $\prob{\nu_{r'+\zeta} < 1} \dotleq \rho^{-d_{ML}(r'+\zeta)}$. Hence
\[\prob{\sigma_{min}(\mathbf{\tilde{R}}_r) \stackrel{.}{<} \rho^{\frac{-\epsilon T}{\kappa}}} \dotleq \rho^{-d_{ML}(r-\epsilon)}\]
for all $r \geq \epsilon >0$.

The association with the singular values
\[ \sigma_1(\mathbf{\tilde{R}}_{r,k})\leq \cdots \leq \sigma_k(\mathbf{\tilde{R}}_{r,k})\]
is made using the interlacing property of singular values of sub-matrices,
which gives that
\begin{align} \label{eq:interlacingProp2}\sigma_i(\mathbf{\tilde{R}}_{r,k})\geq \sigma_i(\mathbf{\tilde{R}}_{r}), \ i\leq k=1,\cdots,\kappa,\end{align}
and for $k=1,\cdots,\kappa$, that
\[\prob{\sigma_{min}(\mathbf{\tilde{R}}_{r,k}) \stackrel{.}{<} \rho^{\frac{-\epsilon T}{\kappa}}} \dotleq \rho^{-d_{ML}(r-\epsilon)}.\]
Finally from the DMT optimality of the exact implementation of the regularized lattice decoder~\cite{GCD:04}, \cite{JE:10}, we have that
\[\prob{\sigma_{min}(\mathbf{\tilde{R}}_{r,k}) \stackrel{.}{<} \rho^{\frac{-\epsilon T}{\kappa}}} \dotleq \rho^{-d_{L}(r-\epsilon)}.\]
This proves Lemma~\ref{lem:lambdamin}.$\square$

%%%%%%%%%%%%%%%%%%%%%%%%%%%%%%%%%%%%%%%%%%%%%%%%%%%%%%%%%%%%%%%%%%%%%%%%%%%%%%%%%%%%%%%%%%%%%%%%%%
%%%%%%%%%%%%%%%%%%%%%%%%%%%%%%%%%%%%%%%%%%%%%%%%%%%%%%%%%%%%%%%%%%%%%%%%%%%%%%%%%%%%%%%%%%%%%%%%%%

\section{Proof for Lemma~\ref{lem:noise_vec}}
\label{app:noise_vec}
For a search radius that grows as $\xi = \sqrt{z \log \rho} \doteq \rho^0$, we first prove that
\[\prob{\|\mathbf{w}^{''}\|^2 > \xi^2} \dotleq \rho^{-z'}\]
for $z>z'>d_L(r)$.
Towards establishing the properties of the equivalent noise $\mathbf{w}^{''}$ (cf.~\eqref{eq:NoiseEquiv}), we consider an equivalent representation of the MMSE-preprocessed lattice decoder and let
(cf.~\cite{WBK:04})
\begin{eqnarray}
\mathbf{Q}\mathbf{R}&=&\left[\begin{array}{l}
\mathbf{Q}_1\\ \mathbf{Q}_2 \\ \end{array}\right]\mathbf{R} =\left[\begin{array}{l}
\mathbf{M}\\ \alpha_r \mathbf{I} \\ \end{array}\right] \in \mathbb{R}^{(n+\kappa) \times \kappa}
\label{eq:ml_mmseqr_equiv}
\end{eqnarray}
be the thin QR factorization of the modified channel matrix, where $\mathbf{Q}_1=\mathbf{R}^{-1}\mathbf{M}\in \mathbb{R}^{n \times \kappa}$, $\mathbf{Q}_2=\alpha_r \mathbf{R}^{-1}\in\mathbb{R}^{\kappa \times \kappa}$ and where $\mathbf{R}^H\mathbf{R}=\mathbf{M}^H\mathbf{M}+\alpha_r^2\mathbf{I}$. It then follows that for $\mathbf{F}=\mathbf{Q}_1^H$, the MMSE-preprocessed lattice decoder is equivalent to lattice decoding in the presence of channel $\mathbf{R}$ and noise \begin{eqnarray} \mathbf{w}^{'}&=&-\alpha_{r}^2\mathbf{R}^{-H}\mathbf{s}+\mathbf{R}^{-H}\mathbf{M}^{H}\mathbf{w} \nonumber \\ &=&-\alpha_{r}\mathbf{Q}_2^{H}\mathbf{s}+\mathbf{Q}_1^{H}\mathbf{w}  \label{eq:noise_qr_equiv} . \end{eqnarray}
Consequently we calculate
\begin{eqnarray}
&& \prob{\|\mathbf{w}^{'}\| > \xi} \nonumber \\
&\leq & \prob{\|-\alpha_{r}\mathbf{Q}_2^{H}\mathbf{s}\|+\|\mathbf{Q}_1^{H}\mathbf{w}\|  > \xi} \nonumber \\
&\stackrel{(a)}{=}& \prob{\|-\alpha_{r}\mathbf{Q}^{H}\left[\begin{array}{l}
\mathbf{s}\\ \mathbf{0} \\ \end{array}\right]\|+\|\mathbf{Q}^{H}\left[\begin{array}{l}
\mathbf{w}\\ \mathbf{0} \\ \end{array}\right]\|  > \xi} \nonumber \\
&{\leq}& \prob{\kappa \bigl(\|\mathbf{w}\| + \sup_{\mathbf{s} \in \mathbb{S}_r^{\kappa}}\|-\alpha_{r}\mathbf{s}\| \bigr) > \xi} \nonumber \\
&\stackrel{(b)}{=}& \prob{\kappa \|\mathbf{w}\| + \kappa K > \xi} \nonumber \\
&{=}& \prob{\kappa\|\mathbf{w}\|  > (z \log \rho)^\frac{1}{2}-\kappa K} \nonumber \\
&\stackrel{(c)}{\leq}& \prob{\kappa\|\mathbf{w}\|  > (z_1 \log \rho)^\frac{1}{2}} \nonumber \\
&{=}& \prob{\|\mathbf{w}\|^2  > \frac{z_1}{\kappa^2} \log \rho} \nonumber \\
&\stackrel{(d)}{=}& \prob{\|\mathbf{w}\|^2  > z_2 \log \rho} \nonumber \\
&\doteq& \rho^{-z_2}
\label{noise_lemma}
\end{eqnarray}
where $(a)$ follows from the MMSE-preprocessed equivalent channel representation (cf.~\eqref{eq:ml_mmseqr_equiv}), and where the inequalities in $(b)$, $(c)$ and $(d)$ follow for some fixed $K$ that upper bounds $\sup_{\mathbf{s} \in \mathbb{S}_r^{\kappa}}\|-\alpha_{r}\mathbf{s}\|$, and for some arbitrary $z_1$, $z_2$ satisfying $z>z_1>z_2>0$ independent of $\rho$. Consequently
\[\prob{\|\mathbf{w}^{''}\| > \xi} = \prob{\|\mathbf{\tilde{Q}}^{H}\mathbf{w}^{'}\| > \xi}\dotleq \rho^{-z'}\]
for some $0 < z' < z_2$, and as a result
\begin{eqnarray*}
\displaystyle\lim_{\rho\to\infty}\frac{\prob{\|\mathbf{w}^{''}\| > \xi}}{\prob{\mathbf{\hat{s}}_{r-ld}\neq \mathbf{s}}} = \displaystyle\lim_{\rho\to\infty}\rho^{(d_L(r)-z')} = 0,
\end{eqnarray*}
where the last equality follows after choosing the search radius such that $z>z' > d_L(r)$. This proves Lemma~\ref{lem:noise_vec}. \ $\square$

%%%%%%%%%%%%%%%%%%%%%%%%%%%%%%%%%%%%%%%%%%%%%%%%%%%%%%%%%%%%%%%%%%%%%%%%%%%%%%%%%%%%%%%%%%%%%%%%%%%%%%%%%
\bibliographystyle{IEEEtran}
%\bibliography{IEEEabrv,../../BIBLIOGRAPHY_AND_DEFS/refs}
\bibliography{IEEEabrv,refs}

\end{document}

%% file: defs.tex
\def\ints{{\mathbb{Z}}}

\def\bfG{{\boldsymbol{G}}}

\def\bfH{{\boldsymbol{H}}}

\def\nt{n_\mathrm{T}}
\def\nr{n_\mathrm{R}}

\def\ints{\mathbb{Z}}

\def\kron{\otimes}
\def\fro{_\mathrm{F}}
\def\kron{\otimes}

\newcommand{\bit}{\begin{itemize}}
\newcommand{\eit}{\end{itemize}}

\newcommand{\bc}{\begin{center}}
\newcommand{\ec}{\end{center}}

\newcommand{\ba}{\begin{array}}
\newcommand{\ea}{\end{array}}

\newcommand{\beq}{\begin{equation}}
\newcommand{\eeq}{\end{equation}}

\newcommand{\beqn}{\begin{equation*}}
\newcommand{\eeqn}{\end{equation*}}

\newcommand{\bean}{\begin{eqnarray*}}
\newcommand{\eean}{\end{eqnarray*}}
\newcommand{\bea}{\begin{eqnarray}}
\newcommand{\eea}{\end{eqnarray}}

\def\C{\mathbb{C}}

\newcommand{\ben}{\begin{enumerate}}
\newcommand{\een}{\end{enumerate}}
\newcommand{\dotleq}{\overset{.}{\leq}}
\newcommand{\dotgeq}{\overset{.}{\geq}}

\newcommand{\defeq}{\triangleq}

\newcommand{\bthm}{\begin{thm}}
\newcommand{\ethm}{\end{thm}}

%% file: macros.tex
\newcommand{\diag}{\mathrm{diag}}

\newcommand{\rank}{\mathrm{rank}}

\def\fro{{\mathrm F}}
\newcommand{\prob}[1]{\mathrm{P}\left(#1\right)}
\newcommand{\expt}[2][]{\mathrm{E}_{#1}\left\{ #2 \right\}}

\newtheorem{theorem}{Theorem}
\newtheorem{corollary}{Corollary}[theorem]
\newtheorem{lemma}{Lemma}